%% file: 0_chiir2024responsible-search.tex
\documentclass[manuscript]{acmart}
\usepackage{booktabs} 
\usepackage{hyperref}
\usepackage{url}
\usepackage{multirow}
\usepackage{cleveref}
\usepackage{float}
\usepackage{balance}
\usepackage{microtype}
\usepackage{comment}
\usepackage{color}
\usepackage{graphics,graphicx}
\usepackage{epstopdf}
\usepackage{colortbl}
\usepackage{diagbox}
\usepackage{tabularx}   
\usepackage[justification=centering]{caption}
\usepackage[inline]{enumitem}

\usepackage{todonotes}
\usepackage{soul}

\newcommand{\post}[0]{\noindent}

\usepackage{tikz}
\usetikzlibrary{arrows}
\usetikzlibrary{decorations.text}

\usepackage{adjustbox}

\newcommand{\figuresize}[1]{\small{#1}}

\usepackage{abbrevs}
\newabbrev\EMCB{\textit{Ethically Minded Consumer Behavior} (EMCB)}[EMCB]
\newabbrev\SRC{\textit{Socially Responsible Consumers} (SRC)}[SRC]

\usepackage[utf8]{inputenc}
\DeclareUnicodeCharacter{2212}{-}

\newcolumntype{D}[1]{>{\centering\arraybackslash}p{#1}}
\newcolumntype{L}[1]{>{\arraybackslash}p{#1}}
\newcolumntype{Y}{>{\centering\arraybackslash}X}

\AtBeginDocument{%
  \providecommand\BibTeX{{%
    \normalfont B\kern-0.5em{\scshape i\kern-0.25em b}\kern-0.8em\TeX}}}

\copyrightyear{2024}
\acmYear{2024}
\setcopyright{acmlicensed}\acmConference[CHIIR '24]{ACM SIGIR Conference on Human Information Interaction and Retrieval}{March}{Sheffield, UK}
\acmBooktitle{ACM SIGIR Conference on Human Information Interaction and Retrieval (CHIIR '24), Sheffield, UK}
\acmPrice{15.00}
\acmDOI{} \acmISBN{}

\makeatletter

\renewcommand\maybe@space@{%
  \maybe@ictrue 
  \expandafter   \@tfor
    \expandafter \reserved@a
    \expandafter :%
    \expandafter =%
                 \nospacelist
                 \do \t@st@ic
  \ifmaybe@ic 
    \space
  \fi
}

\makeatother

\begin{document}

\title{Seeking Socially Responsible Consumers: \\Exploring the Intention-``Search''-Behaviour Gap}

\author{Leif Azzopardi}
\email{leifos@acm.org}
\affiliation{%
  \institution{University of Strathclyde}
  \city{Glasgow}
  \country{UK}
  }
  
\author{Frans van der Sluis}
\email{Frans@hum.ku.dk}
\affiliation{%
  \institution{University of Copenhagen}
  \city{Copenhagen}
  \country{Denmark}
}

\copyrightyear{2024}
\acmYear{2024}
\setcopyright{acmlicensed}\acmConference[CHIIR '24]{Proceedings of the 2024 ACM SIGIR Conference on Human Information Interaction and Retrieval}{March 10--14, 2024}{Sheffield, United Kingdom}
\acmBooktitle{Proceedings of the 2024 ACM SIGIR Conference on Human Information Interaction and Retrieval (CHIIR '24), March 10--14, 2024, Sheffield, United Kingdom}
\acmDOI{10.1145/3627508.3638324}
\acmISBN{979-8-4007-0434-5/24/03}

\begin{abstract}
The increasing prominence of ``\textit{Socially Responsible Consumers}'' has brought about a heightened focus on the ethical, environmental, social, and ideological dimensions influencing product purchasing decisions. 
Despite this emphasis, studies have consistently revealed a significant gap between individuals' intentions to be socially responsible and their actual purchasing behaviors: they often choose products that do not align with their values.
This paper aims to investigate how ``search'' in influences this gap. 
Our investigation involves an online survey of 286 participants, where we inquire about their search behaviors and whether they considered various dimensions—ranging from price and features to environmental, social, and governance issues—in relation to a recent purchase. 
Contrary to expectations of a clear intention-behavior gap, our findings suggest that a considerable number of participants exhibited indifference or lack of information regarding these ``responsible'' aspects.
While, difficulties related to searching for and acquiring information contributed to the gap, including the limited accessibility and reliability of information. 
This suggests that part of the intention-behaviour gap can be framed as an information seeking problem.
Moreover our findings warrant and motivate search systems that help support consumers make more informed and responsible purchasing decisions.
\end{abstract}



\maketitle
\input{1_introduction.tex} 
\input{2_background.tex} 
\input{3_methodology.tex} 
\input{4_results.tex} 
\input{5_discussion.tex} 
\balance

\bibliographystyle{ACM-Reference-Format}
\bibliography{refs.bib,mreferences.bib}
\balance

\clearpage
\setcounter{page}{1}
\appendix
\input{7_supplemental_text.tex} 

\end{document}

%% file: 1_introduction.tex
\section{Introduction}\label{sec_introduction}
Every day millions of consumers search for products online -- trying to find an item that meets their needs and wants at a price that they are willing to pay. 
While price and quality factor heavily into people's purchasing decisions, consumers are increasingly considering other aspects regarding the products they buy~\cite{CarringtonLostGap}. 
This has led to a rise in ethical consumerism and the socially responsible consumer~\cite{Wiederhold2018EthicalIndustry}. 
That is, consumers who want to spend their money on products that align with their values~\cite{Jones2022TheDifference}.
Movements advocating for Fair Trade, Sustainability, Environmentally Friendly, Animal Rights and Welfare, Diversity Equality and Inclusion, Environment Social Governance (ESG) policies, and so on have been raising awareness and bringing consumer's attention towards recognizing other aspects and implications of their purchasing decisions. 
For example, by buying products that have not been tested on animals it avoids issues of animal cruelty, while purchasing products labelled as Fair Trade supports better worker rights. 
The choices consumers make, on aggregate, implicitly support the (mis)behaviours and (mal)practices of 
producers, distributors and sellers of such products~\cite{Jones2022TheDifference}

Being a ``\textit{socially responsible consumer}'', however, is fraught with complexities and challenges.  
And while consumers are becoming or wanting to be more ethically concerned, numerous studies have shown that consumers rarely transform their intentions of being ethical into socially responsible purchases\cite{CarringtonLostGap,Sakib2022EthicalGap,Papaoikonomou2018LookingApproach,Young2010SustainableProducts}. 
This is known as the intention-behaviour gap~\cite{CarringtonLostGap}. 
Many reasons for this gap have been identified including: the higher costs of ethical products, the limited availability of ethical products, and advertising (and false advertising e.g., greenwashing~\cite{deFreitasNetto2020ConceptsReview}) of products.
However, two other reasons often cited include the lack of information available about the different ethical aspects and the difficulties in finding such information.
In this paper, we consider the intention-behaviour gap from a search perspective -- where we aim to investigate people's information seeking behaviors and practices when searching and purchasing products online. 
Our aim is centred on understanding the user's search sided challenges associated with being a socially responsible consumer to identify what difficulties they face when looking for information regarding the ethical aspects of products.


%% file: 2_background.tex
\section{Background} \label{sec_background}
Within Information Retrieval, online shopping has featured as one of the main commercially oriented search tasks on the web~\cite{Broder2002ASearch,Jansen2008DeterminingQueries}. 
This is because search is paramount in getting a good deal~\cite{Ghose2014ExaminingRevenue}. 
This has meant that the transactional web search intent has been a major driver behind the success of large web search engines and e-commerce platforms which benefit from the vast amounts of web advertising revenue and product sales~\cite{Kallumadi2023ECom23:ECommerce}. 
Whether it is the platforms, the retailers or producers, they all have a common goal, to sell (more) products to (more) people to make (more) profit~\cite{Tsagkias2021ChallengesRecommendations}. 
As a result of this lucrative trade, much of the research and the challenges associated with e-commerce and product search has been largely directly towards how companies can advertise and sell their products more efficiently and effectively (e.g., see ACM SIGIR eCom Workshop Series (2017-2023)\cite{Kallumadi2023ECom23:ECommerce} for instance). . 
From a user side, knowing how people search for products and product information has been central in understanding consumer purchasing behaviour. So it is not surprising that researchers (and businesses) have invested heavily in understanding the consumer search process~\cite{Bhatnagar2004OnlineDemographics}.

\subsection{Models of Consumer Buying Process}
Various models and frameworks for how consumers proceed through the purchasing process have been proposed ~\cite{Bloch1986ConsumerFramework,Engel1990CustomerBehavior,RowleyProductPropositions,Punj1983AnMaking,Ke2016SearchProducts,Schmidt1996ASearch}. 
Core to most purchasing decisions is the search for information about the products available, and then evaluating the information about those products to make a purchasing decision~\cite{RowleyProductPropositions}.
According to ~\citet{Ke2016SearchProducts}, during these phases consumers tend to search the market and gather information sequentially on several products. 
Consumers choose which products to gather more information on, and whether to continue to gather information about the products. 
This process may or may not lead to purchasing one of the products, or exiting the market without purchasing. 

Since gathering information is costly, in terms of time and effort~\cite{azzopardi2014economic_models}, consumers often make decisions on incomplete and imperfect information~\cite{WardInternetBranding,Ke2016SearchProducts}. 
The increased availability of information has not alleviated this problem for several reasons: (1) searching involves non-trivial navigation through a large variety of complex web sites --- which can be particularly frustrating and cognitively taxing~\cite{Brynjolfsson1999FrictionlessRetailers}. 
While (2) information asymmetries exist between parties -- where the seller or producer don't disclose or reveal pertinent pieces of information about their products to consumers~\cite{Akerlof1970TheMechanism,Jacoby1974BrandLoad}. 
To this end, branding~\cite{WardInternetBranding} and labelling~\cite{Young2010SustainableProducts,Annunziata2011ConsumersProducts} (e.g. Fair Trade, Energy Efficiency, etc.) provide consumers with information that reduces this asymmetry.
And, (3) with many alternative products to choose from, and each with varying attributes consumers often experience choice and decision overload~\cite{Keller1987EffectsEffectiveness}. 
So rather than helping, more information often decreases consumer's decision making effectiveness~\cite{Keller1987EffectsEffectiveness,Schleenbecker2015InformationCoffee}.  
Consumers accordingly tend to prefer general information over specific information~\cite{Schleenbecker2015InformationCoffee}.

When confronted with ethical information during the search process, consumers often become more concerned and perform more extensive searching~\cite{Zander2012InformationFood}. 
\citet{Schmidt1996ASearch} proposed that consumer information search is mediated by four variables: ability, motivation, costs, and benefits -- the extent to which a consumer will search will therefore depend on their ability to search, how motivated they are in finding out the information that they desire weighted up against the costs and benefits of their searching with respect to their purchasing. 
\citet{Flavian-Blanco2011AnalyzingEngines}, however, argue that it is not just a trade-off between the costs and benefits perceived by consumers, but the search process and the decision making is influenced, consciously or unconsciously, by their affective states and emotional response to the information encountered~\cite{Browne2007CognitiveTasks}. 
As pointed out by \citet{Kuhlthau1991InsidePerspective} the information search process ``\textit{involves the whole experience of the person, feelings as well as thoughts and actions}''. 
The implication being that consumers consider information beyond price and quality when making their purchasing process. 
Of interest in this work is what other aspects consumers search for and how important these different aspects are when making their purchasing decision.

\subsection{Socially Responsible Consumers}\label{bg:src}
With more attention being brought to the ethical aspects of purchasing decision there has been a rise in ethical consumerism and ``\textit{socially responsible}'' consumers~\cite{Sakib2022EthicalGap,Jones2022TheDifference,Casais2022ThePriorities}. 
A ``socially responsible'' consumer, often referred to as a mindful or ethical consumer, is an individual who makes purchasing decisions with a focus on the broader social and environmental impact of their choices, in addition to their personal needs and preferences~\cite{Ellen1991TheBehaviors,CarringtonLostGap,Jones2022TheDifference,Davies2016ConsumerConsumption, Sheth2011MindfulSustainability}. 
These consumers take into consideration a range of ethical, environmental, and social factors when buying products or services~\cite{Jones2022TheDifference,Shaw2006FashionChoice}. 
They prioritize products and services that align with their personal values and ethics~\cite{CarringtonLostGap,Jones2022TheDifference}. 
They include aspects such as human rights, the environment, animal rights, community involvement, and social justice in their purchase decisions~\cite{Jones2022TheDifference}.

Being socially responsible or ethical means different things to different consumers -- so a particular aspect that may be important and relevant to one consumer, may not be to another. 
\citet{Hasanzade2018SelectingShopping} surveyed 249 consumers in Germany and found that most participants were ethically minded consumers (54\%), while the other participants were concerned with price (12\%) and price-quality (34\%). Of those ethically minded, most were concerned about animal rights, followed by labour/human rights, and then environment protection. \citet{Casais2022ThePriorities} surveyed 364 consumers in Portugal, of which most considered themselves as ethical consumers, reported that they were concerns about labour/human rights (31\% ), environmental issues (23\%), animal rights/welfare (17\%), and all three (27\%).
The complexity and variety of ethical concerns translates into highly varied purchasing decisions. 
Not all consumers will address the same issues in identical ways. For instance, a consumer who consciously avoids animal-based products may be indifferent to the manufacturing processes~\cite{Shaw2003EthicsApproach}.

These consumers use their purchasing power as a means to promote positive social and environmental change. 
They believe that by supporting companies and products that align with their values and sustainability goals, they can influence businesses to be more socially and environmentally responsible~\cite{Liu2018TheIntention}. 
This consumer mindset has grown in prominence with increasing awareness of global environmental issues, labor rights, and corporate responsibility, leading to the growth of various certifications and labels to help consumers identify products that meet these criteria~\cite{Jones2022TheDifference}. 
Socially responsible consumers tend to establish an identity rooted in their ethical purchasing decisions, occasionally making personal sacrifices~\cite{Papaoikonomou2018LookingApproach}. 
And, they often tend to communicate their role as advocates for a more sustainable consumption society to others~\cite{Casais2022ThePriorities}.
In this work, we aim to better understand how concerned consumers are about the various ethical dimensions, specifically regarding their search behaviours and experiences when searching for such information when making a purchasing decision.

\subsection{Intention-Behaviour Gap}
The \textit{Theory of Planned Behaviour} seeks to provide an explanation of behaviour by considering how the individual's attitudes and norms influences their intention's and subsequently their behaviour in a casual sequence~\cite{Ajzen1985FromBehavior}.
When the people's behaviours don't align with their intentions, this has been referred to as the \textit{Intention-Behaviour Gap}. Within the literature on consumer purchasing behaviours the misalignment has been of great interest regarding the ethical concerns and observed from many different perspectives~\cite{Moon2004ConsumerHypotheses,lee2010WebUsersCare,Ozcaglar-Toulouse2006InFrance,RowleyProductPropositions, Djafarova2022ExploringBehaviour}. 
For example, 
of 81 self declared green consumers, 30\% reported that they were very concerned about environmental issues but they struggled to translate this into their purchasing decisions~\cite{Young2010SustainableProducts}.

\citet{Uusitalo2004EthicalFinland} conducted a study of Finnish consumers (n=713) and found that while the majority of the participants regarded ethics as important it did not necessarily lead to ethical choices regarding purchases. They found participants were uncertain about which products and which companies were ethical and acting socially responsibly. The major obstacles to being a socially responsible consumer were difficulties in obtaining information, problems in product availability and high prices of ethical products\cite{Uusitalo2004EthicalFinland}.
In a study on sustainable fashion with German participants (n=13), they found that the following barriers impeded ethical purchasing decisions: price, availability, knowledge, transparency, image, inertia and consumption habits~\cite{Wiederhold2018EthicalIndustry}.
And a nine month study of peoples purchasing behaviors ($n=13$), \citet{CarringtonLostGap} found that four interrelated factors affecting the ethical intention-behavior gap: (1) prioritization of ethical concerns; (2) formation of plans/habits; (3) willingness to commit and sacrifice; and (4) modes of shopping behavior~\cite{CarringtonLostGap}. 
Where the major obstacles included alternative personal values, extant habits, inability to form plans, unwillingness to make a commitment/sacrifice, lack of available information and an unwillingness to conduct effortful searches for information, and the distraction of the situational environment in effortful and spontaneous shopping modes.
More recently, \citet{Djafarova2022ExploringBehaviour} interviewed a cohort of participants from Generation Z based in the UK (n=18), and found that they had strong awareness and desire towards ethical and environmental concerns. However, they felt limited by their finances when considering high value items, but tended to exercise more responsibility by recycling, diet choices and reduced consumption~\cite{Djafarova2022ExploringBehaviour}. 
On the other hand, in a study by~\citet{Reczek2018ThatsInformation} they found that undergraduates ($n=236$) would often forget or mis-remember negative ethical information about products they purchased. This was to avoid the emotional distress of having to resolve ethical dilemmas (e.g., these jeans are really nice, but they are made in a sweat shop). 
Following on from these works, in this paper, we consider to what extent the intention-behaviour gap exists and whether this gap is reflected through consumer's information seeking behaviours.




%% file: 3_methodology.tex
\section{Methodology} \label{sec_methodology}

The main research questions addressed in this study are:
\begin{description}
\item[RQ1] To what extent is there a gap between participants' intentions and their online search behaviour related to socially responsible aspects of shopping?
\item[RQ2] What ethical aspects and motivations drive participants' online searches for socially responsible consumerism?
\item[RQ3] What challenges did participants face when conducting online searches to find information about the socially responsible aspects of products?
\end{description}

To examine these research questions, a survey was created and administered. The following subsections provide details on the design (Section \ref{met:design}), survey development (Section \ref{met:survey}), participants (Section \ref{met:participants}), and data analysis (Section \ref{met:analysis}).

\subsection{Design}\label{met:design}
The research employed a mixed method design using a cross-sectional survey to investigate the relationship between information search and responsible consumption behaviors in the context of online product research. 
The survey aimed to gather data on participants' online information seeking practices and their consideration of various product aspects, including price, quality, and responsible consumption dimensions such as environmental and societal impact.

The study was conducted in compliance with our organisation's ethical standards and we received approval from our Ethics Committee (Application No. tba).
Participants' informed consent was obtained, and their privacy and confidentiality were maintained throughout the research process.
The survey was administered online using the Qualtrics survey platform. Participants were informed about the research study and its objectives through a consent form. They were explicitly told about the voluntary nature of participation, and their ability to withdraw from the study at any time without consequences. The survey's estimated completion time was 10-12 minutes. Participants were compensated in line with national wage guidelines.

\subsection{Survey Development}\label{met:survey}
As is customary in the field of consumer research, a survey was created based on related surveys~\citep[e.g.,][]{dastous2008, wang2011ads}.
We adapted questions from existing surveys and introduced new ones to suit the specific study context.
Differing from conventional surveys, we encouraged open-ended responses to elicit richer, context-specific data from participants. 
These adaptations served to capture nuanced aspects of online product searches in responsible consumer behavior. 
We piloted the survey to fine tune the questions (this participants were not included in our analysis).
The full survey, excluding the informed consent questions, can be found in Supplemental Appendix~\ref{app:questions}.
The questionnaire was divided into three parts:

\subsubsection*{Part 1: Recent Purchase Questions}
Part 1 asked participants to describe a concrete recent purchase decision.
By focusing on tangible, real-world purchase rather than a hypothetical scenario, the study aimed to mitigate potential social desirability bias. This approach encouraged participants to reflect on concrete experiences rather than speculative situations.
The first question asked participants for a high-level description of the product they had purchased, such as a phone, ice skates, skirt/shorts, or any other product type.
The next question asked which category the purchased product, was within, such as clothing, household appliances, consumer electronics, books, sports equipment, DIY tools, and other categories. These categories were sourced from Statista.com\footnote{\url{https://www.statista.com/statistics/276846/reach-of-top-online-retail-categories-worldwide/}, last accessed on \today}.
We also included questions to inquire about the duration and intensity of the participants' search process. These questions aimed to assess the significance of their reported purchase. In the creation of these questions, care was taken to adhere to survey guidelines, ensuring that the questions requested concrete numerical responses, avoided ambiguity, and provided a balanced set of answer options \citep[][ch. 11]{clark2021bryman}.

\subsubsection*{Part 2: Individual aspect questions}
Part II of the survey contained questions regarding individual aspects. A set of five aspects was randomly sampled from the list of thirteen aspects (see Table \ref{tab:aspects} in Appendix \ref{app:aspects}), presented one at a time.
For each aspect present, a brief explanation was shown (see Table \ref{tab:aspects}, column Elaboration) to contextual the forthcoming questions.
These aspects were gathered from previous surveys on responsible and ethical consumer to ensure their relevance (see Section \ref{bg:src}).

The first set of questions (\ref{q:importance} -- \ref{q:searched}) was designed to evaluate the relevance of each aspect to their purchasing decision.
The questions evaluated their perceived importance of an aspect (\ref{q:importance}), their intention to consider it during their decision-making (\ref{q:considered}), and finally, whether they actively searched for information for that aspect (\ref{q:searched}).

For participants who indicated that they hadn't searched for information about a particular aspect, were presented with follow-up questions (\ref{q:reasonsno}). These included predefined answer options designed to encompass common reasons for not searching, along with an ``Other'' option and a free-text box that they could use to explain why they didn't search for information about this aspect.

Participants who indicated that they had searched for a specific aspect were presented with four sets of three questions designed to explore the challenges they encountered during their search. These questions were structured around Brehm's model of task performance, which emphasizes four key factors potentially impeding engagement: success expectancies, perceived value, effort, and perceived difficulty (\ref{q:effort} -- \ref{q:difficulty}). Furthermore, participants were encouraged to offer an open-ended response, sharing their motivations and insights gained during their search for information on the aspect (\ref{q:reasonsyes}).

\newcommand{\mylistlabelfont}[1]{{\normalfont\itshape\space #1 --}}
\newlist{inlinedesc}{description*}{1}
\setlist*[inlinedesc,1]{%
  font=\mylistlabelfont, mode=unboxed
}

Brehm's model has been used extensively in educational contexts but not in the context of retrospective evaluations of online searches.
The model describes the importance of:
\begin{description}
 \item[Success expectancies] likelihood of achieving favorable outcomes, such as finding valuable information during the search, based on their actions;
 \item[Perceived value] importance or utility of the task, reflecting how valuable the information sought would have been for their purchasing decision;
 \item[Effort] level of exertion or resources dedicated to complete the task successfully, and;
 \item[Perceived difficulty] ease or difficulty of finding relevant information during their search on the chosen aspect.
\end{description}
The questions in our survey were based on surveys that had previously utilized Brehm's factors within educational settings. For instance, ``\textit{What I learned in today’s class was useful}'' \citep[][p. 1126]{tanaka2014} was rephrased for our retrospective search scenario as ``\textit{I learned information about "[Aspect Name]" that was useful in making my purchasing decision}'' (\ref{q:value}.3).

Together, these questions aimed to uncover why and how participants navigated their searches for aspects related to responsible consumer behavior.
By combining both quantitative and qualitative questions,
Part 2 was designed to provide a comprehensive understanding on both motivations and challenges.

\subsubsection*{Part 3: Additional Insights and EMCB Scale}

Part 3 of the survey aimed to gather additional insights into participants' responsible consumer behavior. This section included open-ended questions that encouraged participants to share other criteria they considered in their purchasing decisions, additional information they searched for, and any other challenges they generally face during online product searches. Following these questions, participants completed the standardized \EMCB scale~\citep{SUDBURYRILEY20162697}, providing a quantitative assessment of their eco-friendly consumption practices. This scale is extensively tested among consumers in multiple countries, demonstrating its cross-cultural validity.

\begin{figure*}[t!]
\makebox[\textwidth][c]{\input{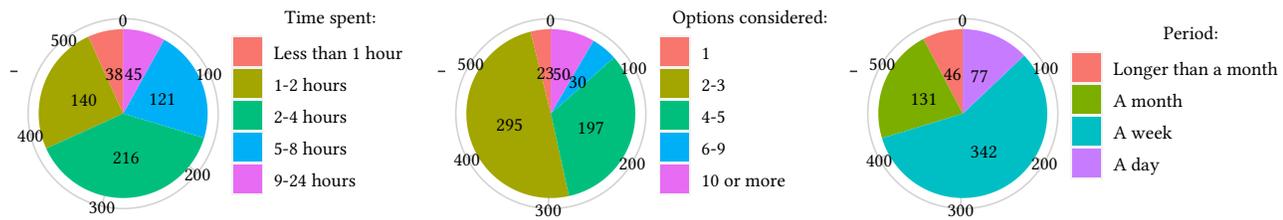}}
\vspace{-5mm}
\caption{Pie charts showing the time invested, options considered, and period covered by the reported purchase decisions.}\label{fig:piechart}
\end{figure*}

\subsection{Participants}\label{met:participants}
The study involved 286 participants who were recruited through the online platform Prolific. Eligible participants were required to be 18 years or older, capable of giving informed consent, and must have made a recent online purchase valued at approx. \$100/€100 or more within the past 3 months, where they compared multiple products before making their final decision.
Participants were of diverse backgrounds and geographic locations, contributing to the study's generalizability.

In total, 297 individuals entered the survey, reflecting on 1698 aspects of their purchases. Of those who entered, six participants didn't give their consent, two didn't meet the purchase requirement, and an additional three didn't complete the survey, resulting in 286 participants.

Of the $286$ participants who fully completed the study, $99$ were female, $187$ were male. Their ages ranged from $18$ to $75$, with most participants in the $26-35$ age bracket ($35.90\%$). Other age brackets and percentages included: $18-25$ ($26.10\%$); $36-45$ ($19.90\%$); the remainder were over $45$.
Participants resided in predominantly European and North-American countries, with the addition of South-Africa as other major place of residence. The top-5 countries of residence were: United Kingdom (97), Poland (41), United States (34), and South Africa (24). 

\subsection{Data Analysis and Reporting}\label{met:analysis}
Before proceeding with the analysis, the collected data underwent several pre-processing steps to ensure data quality. These steps included the removal of participants who did not meet the specified criteria and ensuring that no personally identifiable information was associated with the responses. Additionally, questions that were reversed scaled were re-scaled.

For the scales consisting of multiple questions, Cronbach's alpha was computed. The Cronbach's alpha values for each scale were as follows: \EMCB ($\alpha = 0.94$), effort ($\alpha = 0.38$), difficulty ($\alpha = 0.79$), perceived value ($\alpha = 0.82$), and success ($\alpha = 0.70$).
With the exception of the effort scale, these values were all satisfactory.
The effort scale was accordingly excluded from further analysis.

Aspects were categorized into five themes to facilitate concise reporting. These themes were formed deductively following related work (see \ref{bg:src}) and further refined inductively by examining participants' textual responses explaining their reasons for searching. This process resulted in the five themes shown in Table \ref{tab:aspects}.
Of the five themes, ``Product Evaluation'' and ``Reputation and Governance'' will not be reported on as they do not capture ethical considerations.
%
Furthermore, participants' free-text reasons for not searching (\ref{q:reasonsno}) were grouped thematically by Author 2, creating the structure of Section \ref{res:challenges}. Any disagreements or uncertainties in labeling were discussed and revised, following the group-labels' review by Author 1. This dual-author approach to label validation was employed to enhance the reliability and accuracy of the thematic coding

Data collected from the survey underwent a series of analyses aimed at addressing the research questions. Descriptive statistics, including frequencies and means, provided an overview of the data. Inferential statistical techniques using R were employed, including linear regression, correlational analysis, and the computation of confidence intervals, to test for relations. The following section presents the results of these data analyses.

%% file: 4_results.tex
\section{Results} \label{sec_results}
Participants reported on overall fairly extensive purchase decisions. Figure \ref{fig:piechart} shows the majority of participants spent between 2 and 4 hours, considered 2 to 3 options, and completed their purchases over the course of a week.
There is nevertheless considerable spread in the answers, with some participants reporting on purchase decisions that lasted for over a month, took up to 24 hours, and compared 10 or more options. 
Purchases that took less than 1 hour and considered only one option were uncommon in the answers.
These descriptive findings underscore that participants described purchase decisions that demanded a substantial level of effort as well as information to complete.

Following will be an approximation of the extent of a gap between intention and searches on responsible consumption aspects (Section \ref{res:gap}),
a description of the role of search in the evaluation of ethical aspects (Section \ref{res:search}),
and an examination of common challenges perceived with these searches (Section \ref{res:challenges}).

\subsection{Intention-Behaviour Gap}\label{res:gap}
For the participant's reported purchase,
our expectation was that intentions towards being socially responsible would translate into search activity for those aspects. This expectation will be explored quantitatively through a comparison of importance ratings, and whether participant's considered and then searched for such aspects, and in light of their \EMCB scores.

\begin{figure*}[h!]
\makebox[\textwidth][c]{\input{./objects/importance_graph.tex }}
\vspace{-.5cm}
\caption{Importance values including $95\%$ confidence intervals for each of the aspects surveyed. Values correspond to (1) not at all, (2) slightly, (3) moderately, (4) very, and (5) extremely important. 
}\label{fig:importance}
\end{figure*}
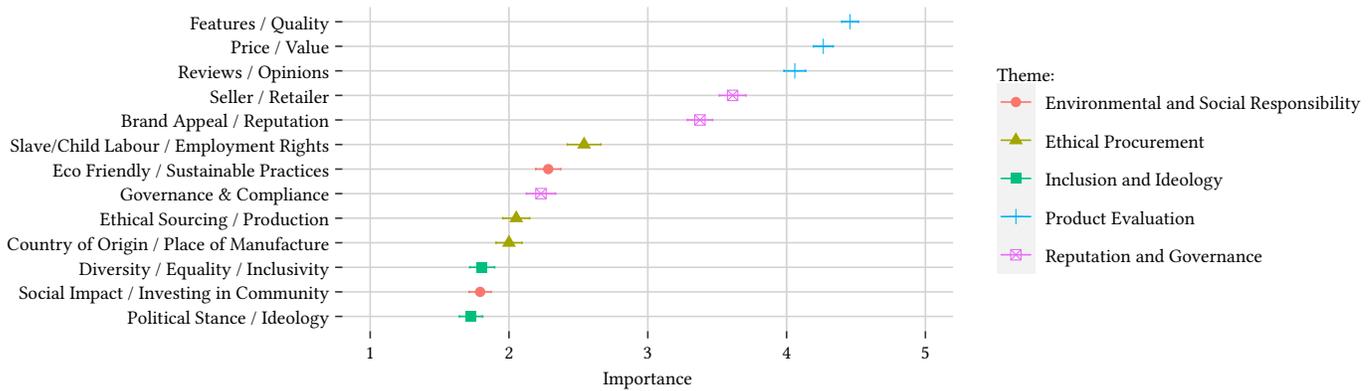

Participants rated the importance of in total thirteen aspects to their recent purchase decision.
Figure \ref{fig:importance} shows that, overall, aspects related to product evaluation held the highest importance.
This aligns with our expectations, indicating that purchase decisions primarily revolve around product quality, features in relation to price, with reviews and opinions serving as vital sources of information.
Furthermore, aspects concerning the seller/retailer and brand appeal/reputation were also considered relatively significant, likely because they contribute to assessing the product's quality and the credibility of the information available.

Contrary to our expectations, aspects associated with ethical, social, and governance considerations were perceived as relatively unimportant in influencing participants' purchase decisions.
This limited attention for secondary aspects, not directly related to product and retailer evaluation, is in line with a mean of $2.87$ on the \EMCB scale (range 1--5).
It suggests that responsible consumerism was not a strong priority for our participants and that related aspects were of less importance to the recent purchase decision they reported on. While this may appear to contrast past work, many of the previous studies sampled consumers ethically inclined.
This perspective was not universal. After performing an exploratory factor analysis, we found that there was three distinct clusters where 23.8\% were highly ethical inclined with \EMCB scores ($>3.8$), while 28.5\% were largely indifferent with low \EMCB scores ($<2.1$), and the remaining 47.6\% were somewhat ethical with average \EMCB that varied. For this middle group participants were high in some dimensions but low in others, and this was different for different participants, confirming that people have a variety and mixture of concerns (as shown in \cite{Hasanzade2018SelectingShopping}).

\begin{figure*}
\makebox[\textwidth][c]{\input{./objects/searches_chart.tex }}
\vspace{-.5cm}
\caption{Stacked bar plot showing, per theme, the percentage of participants that considered and searched for it. Themes are ranked based on their considered, searched percentage.}\label{fig:searches}
\end{figure*}
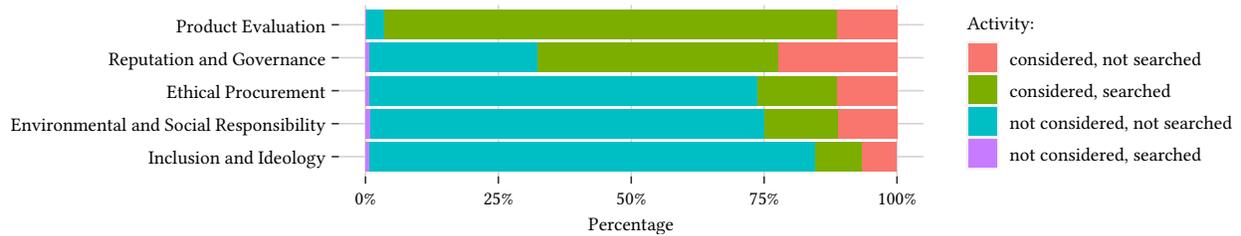

Figure \ref{fig:searches} shows a breakdown of considerations and search activity per theme.
The distribution of themes that participants considered and actively searched for during their purchase decision corresponds to the importance value reported in Figure \ref{fig:importance}.
Notably, participants predominantly considered and searched for aspects related to product evaluation, while relatively few considered potential environmental, social and ethical issues or aspects related to inclusion and ideology.

In Figure \ref{fig:searches}, between $4.09\%$ and $23.72\%$ of participants considered an aspect, but did not search for it.
This indicates that aspects, even when considered, do not necessarily translate into search activity.
In fact, from those participants that considered an aspect, the percentage that also searched drops from $89.03\%$ for product evaluation to $51.85\%$ for environmental and social responsibility.
This signals a growing gap between intention and accompanying information acquisition.
Participants were much less likely to have searched for aspects more distant from product evaluation, even when they considered them.

Overall, the results on importance, considered, and searched ratings suggest a widening gap between intentions and actions at various stages of the decision-making process.
This expanding gap became further apparent when examining the relationship between participants' \EMCB scores and aspects related to the environmental and social responsibility theme.
The \EMCB explained in total $R^2=26.11\%$ of variance for this theme, leaving a noteworthy proportion of variance unexplained.
The correlation furthermore decreased from a strong correlation with importance ratings at $r=.511$ ($t(241)=8.96, p<.001$) to a moderate correlation with considerations ($r=.346, t(244)=5.76, p<.001$) and weakens further when assessing search actions ($r=.300, t(244)=4.92, p<.001$).
%
%
%
This progressive weakening of the correlation, together with the proportion of variance left unexplained, highlights the challenge of translating environmental and social intentions into concrete actions during purchase decisions.

\subsection{The Role of Search}\label{res:search}
It is well-known that search plays a pivotal role in purchase decisions (see Section~\ref{sec_background}). To explore this role, particularly in relation to socially responsible aspects, participants' reasons for searching will be reviewed. 
This includes both aspect importance scores (see Figure \ref{fig:importance}) and 320 free-text reasons (\ref{q:reasonsyes}).

\subsubsection{Ethical Procurement}

Aspects pertaining to ethical procurement were deemed as at least slightly important by nearly half of the participants ($48.31\%$).
They were mentioned in $7.75\%$ of comments and were
interpreted these aspects both in relation to product evaluation, and in relation to broader considerations concerning labor and work conditions:
\begin{center}
\begin{tabular}{p{1.0\linewidth}}
\midrule
\post ``\textit{I wanted to find the country of origin because some countries make better and more reliable products.}'' (Country of Origin / Place of Manufacture - Participant 37) \\ 
\post ``\textit{I wanted to learn about the ethical sourcing and production practices of a product to make an informed purchasing decision (...)}'' (Ethical Sourcing / Production - Participant 195) \\ 
\post ``\textit{I wanted to make sure that the clothes that i was buying were not manufactured under slave conditions or by children.(...)}'' (Slave/Child Labour / Employment Rights - Participant 224) \\ 
\midrule
\multicolumn{1}{r}{Production and Origin - Reasons to search}
\end{tabular}
\end{center}

The first comment was shared by several other comments, as participants generally considered country of origin as a proxy for product quality. This viewpoint was of primary concern for participants within this theme and contributed directly to their product evaluation.
%
%
In contrast, the second and third comment adopted a more explicit ethical stance.
These and other comments within this theme mentioned a range of considerations encompassing labor conditions, fair trade practices, sustainable sourcing, and included detailed factors such as certifications and materials sourcing.
%
These participants demonstrated a heightened awareness of the ethical implications of their purchasing choices.
%



\subsubsection{Environmental and Social Responsibility}
Aspects pertaining to environmental and social responsibility were deemed as at least slightly important by nearly half of the participants ($49.13\%$).
%
With only $2.82\%$ of mentions, few participants gave reasons to search for aspects on environmental and social responsibility.
Our participants interpreted these aspects mostly in relation to the specific products they were purchasing, occasionally extending their considerations to the supplier's practices.
%
From the comments, few were raised on social impact, while many were on environmental impact:
\begin{center}
\begin{tabular}{p{1.0\linewidth}}
\midrule
\post ``\textit{Interested in: Community contributions, philanthropy, inclusivity efforts. Impact on decision by supporting brands aligning with values, positively impacting society.}'' (Social Impact / Investing in Community - Participant 164) \\ 
\post ``\textit{Whether the product was ethically manufactured and if it could be recycled at the end of its life}'' (Eco Friendly / Sustainable Practices - Participant 249) \\ 
\midrule
\multicolumn{1}{r}{Environmental and Social Responsibility - Reasons to search}
\end{tabular}
\end{center}
While the first comment displayed a broader interest in the societal contributes of a supplier, this level of engagement was an exception within our data set.
The second comment was shared by more participants, and often through a specific and detailed exploration of environmental facets.
Our participants have sought information on energy labels, product life expectancy, packaging, product recycling, and more. These comments show an overall high level of knowledge and engagement with the theme.
%

\subsubsection{Inclusion and Ideology}
Aspects pertaining to inclusion and ideology were considered of no concern to their purchase decisions by most participants ($63.04\%$). Only a few participants ($2.82\%$ of mentions) provided reasons in favor of searching on this theme:
\begin{center}
\begin{tabular}{p{1.0\linewidth}}
\midrule
\post ``\textit{Did that brand still sell in russia. Because I don't want to support them financially.}'' (Political Stance / Ideology - Participant 78) \\ 
\post ``\textit{I always want to know those details when I am ready to buy something. I want to know the country of production and the trade mark and to check if there are problems regarding diversity /equality and inclusivity.}'' (Diversity / Equality / Inclusivity - Participant 127) \\ 
\midrule
\multicolumn{1}{r}{Inclusion and Ideology - Reasons to search}
\end{tabular}
\end{center}
These comments described participants searching for the values a company represents and trying to match those values to their own.
Such inquiries extended beyond mere compliance with labor and environmental regulations as in the ethical procurement theme and instead sought companies actively pursuing higher ethical goals.

%
%

Across themes, participants' reasons for searching showed that their interpretation followed the intended meaning of the aspects.
They furthermore highlighted that participants needed extensive prior knowledge
to identify the key features or aspects associated with high quality, safe, ethically procured, and environmentally and socially responsible products.
This necessary knowledge encompassed domain-specific information, such as insights into supplier reputations, certification standards, recycling options, and other intricate details.

\subsection{Perceived Search Challenges}\label{res:challenges}
So far, the data revealed common reasons why users choose to search for specific aspects, as well a progressive gap from intentions to seeking. 
This section delves deeper into the underlying reasons for this gap. We explore the challenges participants faced and expected to face while searching for specific aspects.

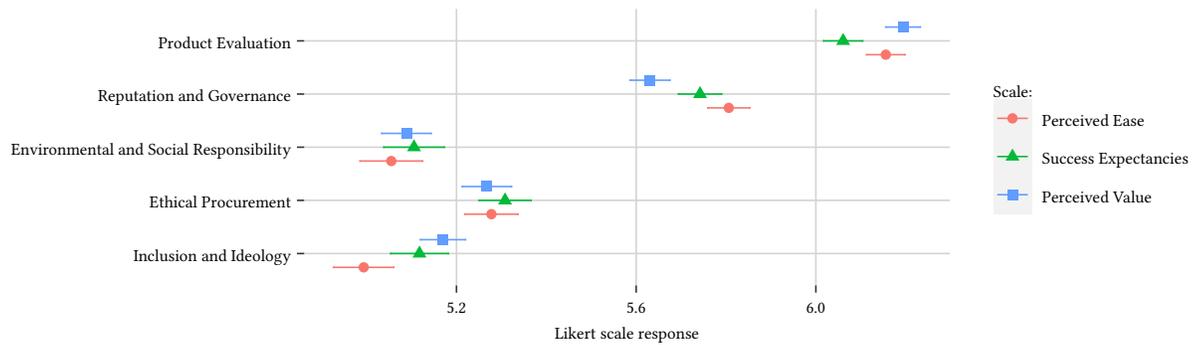
\begin{figure*}
\figuresize{\input{./objects/brehm_chart.tex }}
\vspace{-.5cm}
\caption{Likert scale responses for the ease, perceived value, and success expectancies for searches. Non-adjusted $95\%$ confidence intervals are included as error bars. }\label{fig:brehm}
\end{figure*}

The scales derived from Brehm's model give a first quantitative approximation of the challenges that arise while people try to search for secondary aspects.
Figure \ref{fig:brehm} shows that, overall, participants' perceived ease of use, perceived value, and success expectancies for their searches were lower for secondary than for primary aspects. 
The three factors visually co-varied with each other, which can be expected as all three factors co-determine task effort following Brehm's model of task motivation.
ANOVAs confirm a significant difference for all three scales over themes with $F(4, 599) = 20.29, p < .001$ for ease of access, $F(4, 599) = 24.35, p < .001$ for perceived value, and $F(4, 599) = 14.58, p < .001$ for success expectancies.
%
Pairwise comparisons after Bonferonni correction confirmed differences between primary and secondary themes for all three scales, but not within secondary themes and with the following exceptions: Environmental \& social responsibility -- reputation \& governance for success expectancies; inclusion \& ideology -- reputation \& governance for perceived value, and; ethical procurement -- reputation \& governance for both.
%
These results show that users encountered significantly more difficulties and held lower success expectancies for searches on secondary themes than product evaluation and, to a lesser extent, reputation and governance. 

%
We will further analyze the challenges highlighted by Brehm's scales by investigating the reasons users provided for searching for the listed themes. 
This investigation covers both 1578 selections of pre-listed reasons and 48 free-text responses for \ref{q:reasonsno}. Two pre-listed reasons, '\textit{I can't do anything about it}' (149 selections) and '\textit{There are no good alternatives for this aspect, so I don't look for it},' (93 selections) are excluded as they do not pertain to search-related issues.
We extracted common difficulties and obstacles that users encountered from participants' reasons, organizing this section around these shared challenges.


\subsubsection{Lack of relevance}
Participants commented ($50\%$ of comments) on the perceived lack of relevance of secondary aspects to their purchasing decisions.
Some of the participants voiced disagreement with an aspect altogether:
\begin{center}
\begin{tabular}{p{1.0\linewidth}}
\midrule 
\post ``\textit{A company that is in business to make money should only concern itself with making the best product it can, ethically - and therefore to maximise profits. It should not concern itself with cultural/political ideologies.}'' (Political Stance / Ideology - Participant 242) \\ 
\post ``\textit{I hope they get rid of Diversity / Equality / Inclusivity, imagine having this in the airplane/pilot sector, I want to make sure I'm on a plane that is safe not on a plane to where the pilot can't fly it}'' (Diversity / Equality / Inclusivity - Participant 98) \\ 
\post ``\textit{I'm against it}'' (Social Impact / Investing in Community - Participant 103, C1) \\ 
\post ``\textit{I would rather purchase something that is of good value to me immediately, eco friendly/sustainability is not a concern for me.}'' (Eco Friendly / Sustainable Practices - Participant 162) \\ 
\midrule
\multicolumn{1}{r}{Lack of perceived relevance of an aspect}
\end{tabular}
\end{center}
These comments one by one indicated they did not perceive importance in the aspects we listed. These honest replies concur with the low importance ratings observed in Figure \ref{fig:importance} and predominantly occurred for the ideology \& inclusion and environmental \& social responsibility themes.
They were shared with 202 selections of '\textit{I don't care about it}'.
Others voiced similar concerns, not about an aspect in general, but in relation to a particular product they were buying:
\begin{center}
\begin{tabular}{p{1.0\linewidth}}
\midrule 
\post ``\textit{It had nothing to do with the product}'' (Social Impact / Investing in Community - Participant 8) \\ 
\post ``\textit{I wouldn't think that an oven would be related. However, to note I do consider it when buying clothing.}'' (Slave/Child Labour / Employment Rights - Participant 42) \\ 
\post ``\textit{With gaming consoles I don't consider it but maybe with clothing I would}'' (Eco Friendly / Sustainable Practices - Participant 248) \\ 
\midrule
\multicolumn{1}{r}{Lack of perceived relevance to a product}
\end{tabular}
\end{center}
These comments represent a selection from a broader range of participant responses, all of which underscore a perceived lack of relevance of certain aspects to
a purchasing decision.
This perceived irrelevance aligns with 294 selections of '\textit{It would not have made a difference to my purchasing decision}'.
These more nuanced comments suggest that participants gauge the importance of these aspects based on their awareness of product-specific risks and considerations.
%

\subsubsection{Limited accessibility}
In $14.58\%$ of comments,
participants expressed concerns regarding the challenges they faced in accessing or locating information related to specific aspects:
\begin{center}
\begin{tabular}{p{1.0\linewidth}}
\midrule 
\post ``\textit{Virtually impossible to assess this in my opinion, there simply isn't the information there.}'' (Ethical Sourcing / Production - Participant 277) \\ 
\post ``\textit{I often can't find the information I'm looking for.}'' (Diversity and Inclusion / Diversity / Equality / Inclusivity - Participant 1902) \\ 
\post ``\textit{I wasn't sure where to find this information.}'' (Ethical Procurement / Slave/Child Labour / Employment Rights - Participant 1246) \\ 
\midrule
\multicolumn{1}{r}{Lack of accessibility}
\end{tabular}
\end{center}
These comments collectively highlight the challenges participants expected when attempting to access or locate information on secondary aspects. Participants conveyed that these search difficulties were primarily attributable to either the unavailability or lack of findability of relevant information.
These comments are echoed in 122 selections of '\textit{I can never find information about this aspect, so I don't look for it}'.
As exemplified by the following comments, these challenges necessitated considerable effort to overcome:
\begin{center}
\begin{tabular}{p{1.0\linewidth}}
\midrule 
\post ``\textit{The information isn't readily available, but some of the alternative products I was considering came from areas where there was a high probability of child/slave labor being used. After checking for the manufacturing areas, I removed those products from consideration.}'' (Slave/Child Labour / Employment Rights - Participant 280) \\ 
\post ``\textit{If the information was on the product description, I would consider it, but otherwise, it was a topic that I did not check.}'' (Political Stance / Ideology - Participant 1984) \\ 
\midrule
\multicolumn{1}{r}{Accessibility and decision-making processes}
\end{tabular}
\end{center}
These two comments exemplify the challenges faced in accessing information.
They are supported by 157 selections of '\textit{It is too time consuming to find information about this aspect}'.
%
This underscores the effects of limited accessibility on participants' ability and willingness to consider secondary aspects in their decision-making process, partly explaining why these aspects were not more considered (see Figure \ref{fig:searches}).



\subsubsection{Prior knowledge}
Participants highlighted the influence of prior knowledge on their decision-making processes in $14.58$ of comments:
%
\begin{center}
\begin{tabular}{p{1.0\linewidth}}
\midrule 
\post ``\textit{Have used before, so already know}'' (Seller / Retailer - Participant 58) \\ 
\post ``\textit{I mostly considered well-known companies, that I would assume are reputable.}'' (Governance \& Compliance - Participant 218) \\ 
\midrule
\multicolumn{1}{r}{Existing knowledge rendering search unnecessary}
\end{tabular}
\end{center}
These comments emphasize that participants relied on their existing knowledge, leading them to conclude that further search was unwarranted.
They are confirmed by 199 selections of '\textit{I already knew the information I needed about this aspect}'.
Conversely, some participants lacked relevant knowledge, which made search challenging or less apparent:
\begin{center}
\begin{tabular}{p{1.0\linewidth}}
\midrule
\post ``\textit{I don't know anything about governing and compliance, that is why I did not check.}'' (Governance \& Compliance - Participant 147) \\ 
\post ``\textit{It didn’t occur to me as none of the products were advertising their green credentials.}'' (Social Impact / Investing in Community - Participant 25) \\ 
\post ``\textit{Same case as the previous one, it did not occur to me to inform myself about this topic :(}'' (Diversity and Inclusion / Diversity / Equality / Inclusivity - Participant 222) \\ 
\midrule
\multicolumn{1}{r}{Lack of relevant knowledge impeding search}
\end{tabular}
\end{center}
These comments illustrate instances where participants lacked the necessary knowledge to engage in search or even consider it. 
%
It aligns with the most-selectioned pre-listed reason of '\textit{It is not something that I have considered before}' (362 selections).
This underscores how participants' prior knowledge and awareness, or the lack thereof, can play a pivotal role in shaping their decision-making processes.
And points to a knowledge gap, rather than an intention gap.


\subsubsection{Reliability concerns}
In $12.5\%$ of comments, participants  raised concerns regarding the reliability of information sources.
These concerns were not included the pre-listed reasons, and only originate from participant-supplied comments.
Their comments either reflected skepticism about reviews and sources, or challenges related to assessing specific aspects:
\begin{center}
\begin{tabular}{p{1.0\linewidth}}
\midrule 
\post ``\textit{I am skeptical about how genuine the reviews I find online are}'' (Reviews / Opinion - Participant 50) \\ 
\post ``\textit{It's hard to find reliable sources (Eco Friendly / Sustainable Practices - Participant 265)} \\ 
\post ``\textit{I don't think most of that information is actually accurate.}'' (Diversity / Equality / Inclusivity - Participant 59) \\ 
\midrule
\multicolumn{1}{r}{Skepticism about Reviews and Information Sources}
\end{tabular}
\end{center}
The comments suggest a lack of trust in the authenticity and accuracy of online reviews and sources, particularly when evaluating aspects related to eco-friendliness, sustainability, diversity, and inclusivity. This skepticism potentially hinders participants from relying on such information in their decision-making process, as illustrated by the following comments:
%
\begin{center}
\begin{tabular}{p{1.0\linewidth}}
\midrule 
\post ``\textit{Most items are either manufactured in China and other parts of Asia or the parts are.  The listed manufacturing country is only the final assembly in a lot of cases, so it means little.}'' (Country of Origin / Place of Manufacture - Participant 277) \\ 
\post ``\textit{(...) It is much easier to assess the ethics of retailers I feel but quite difficult to gain much information about manufacturing companies.  What impartial sources could I use to find this information?  There is a lot of misinformation on the web, and it can be very difficult to find a balanced view of a company. (...)}'' (Inclusion and Ideology / Political Stance / Ideology - Participant 277) \\ 
\midrule
\multicolumn{1}{r}{Challenges in Assessing Aspects}
\end{tabular}
\end{center}
These comments reflect participants' challenges in assessing specific aspects related to their purchase decisions, notably the country of origin and the ethical stance of manufacturing companies. 
The collective concerns presented in both sets of comments highlight the critical role that information reliability plays in participants' decision-making processes. 



%% file: objects/importance_graph.tex
\begin{tikzpicture}[x=1pt,y=1pt]
\definecolor{fillColor}{RGB}{255,255,255}
\path[use as bounding box,fill=fillColor,fill opacity=0.00] (0,0) rectangle (523.96,144.54);
\begin{scope}
\path[clip] (  0.00,  0.00) rectangle (523.96,144.54);

\path[] (  0.00,  0.00) rectangle (523.96,144.54);
\end{scope}
\begin{scope}
\path[clip] (131.53, 21.83) rectangle (362.54,144.54);

\path[] (131.53, 21.83) rectangle (362.54,144.54);
\definecolor{drawColor}{RGB}{211,211,211}

\path[draw=drawColor,line width= 0.6pt,line join=round] (131.53, 27.41) --
	(362.54, 27.41);

\path[draw=drawColor,line width= 0.6pt,line join=round] (131.53, 36.70) --
	(362.54, 36.70);

\path[draw=drawColor,line width= 0.6pt,line join=round] (131.53, 46.00) --
	(362.54, 46.00);

\path[draw=drawColor,line width= 0.6pt,line join=round] (131.53, 55.30) --
	(362.54, 55.30);

\path[draw=drawColor,line width= 0.6pt,line join=round] (131.53, 64.59) --
	(362.54, 64.59);

\path[draw=drawColor,line width= 0.6pt,line join=round] (131.53, 73.89) --
	(362.54, 73.89);

\path[draw=drawColor,line width= 0.6pt,line join=round] (131.53, 83.18) --
	(362.54, 83.18);

\path[draw=drawColor,line width= 0.6pt,line join=round] (131.53, 92.48) --
	(362.54, 92.48);

\path[draw=drawColor,line width= 0.6pt,line join=round] (131.53,101.78) --
	(362.54,101.78);

\path[draw=drawColor,line width= 0.6pt,line join=round] (131.53,111.07) --
	(362.54,111.07);

\path[draw=drawColor,line width= 0.6pt,line join=round] (131.53,120.37) --
	(362.54,120.37);

\path[draw=drawColor,line width= 0.6pt,line join=round] (131.53,129.67) --
	(362.54,129.67);

\path[draw=drawColor,line width= 0.6pt,line join=round] (131.53,138.96) --
	(362.54,138.96);

\path[draw=drawColor,line width= 0.6pt,line join=round] (142.03, 21.83) --
	(142.03,144.54);

\path[draw=drawColor,line width= 0.6pt,line join=round] (194.53, 21.83) --
	(194.53,144.54);

\path[draw=drawColor,line width= 0.6pt,line join=round] (247.04, 21.83) --
	(247.04,144.54);

\path[draw=drawColor,line width= 0.6pt,line join=round] (299.54, 21.83) --
	(299.54,144.54);

\path[draw=drawColor,line width= 0.6pt,line join=round] (352.04, 21.83) --
	(352.04,144.54);
\definecolor{fillColor}{RGB}{0,191,125}

\path[fill=fillColor] (178.20, 25.45) --
	(182.12, 25.45) --
	(182.12, 29.37) --
	(178.20, 29.37) --
	cycle;
\definecolor{fillColor}{RGB}{248,118,109}

\path[fill=fillColor] (183.62, 36.70) circle (  1.96);
\definecolor{fillColor}{RGB}{0,191,125}

\path[fill=fillColor] (182.43, 44.04) --
	(186.35, 44.04) --
	(186.35, 47.96) --
	(182.43, 47.96) --
	cycle;
\definecolor{fillColor}{RGB}{163,165,0}

\path[fill=fillColor] (194.53, 58.35) --
	(197.18, 53.77) --
	(191.89, 53.77) --
	cycle;

\path[fill=fillColor] (197.28, 67.64) --
	(199.92, 63.07) --
	(194.64, 63.07) --
	cycle;
\definecolor{drawColor}{RGB}{231,107,243}

\path[draw=drawColor,line width= 0.4pt,line join=round,line cap=round] (204.67, 71.93) rectangle (208.60, 75.85);

\path[draw=drawColor,line width= 0.4pt,line join=round,line cap=round] (204.67, 71.93) -- (208.60, 75.85);

\path[draw=drawColor,line width= 0.4pt,line join=round,line cap=round] (204.67, 75.85) -- (208.60, 71.93);
\definecolor{fillColor}{RGB}{248,118,109}

\path[fill=fillColor] (209.36, 83.18) circle (  1.96);
\definecolor{fillColor}{RGB}{163,165,0}

\path[fill=fillColor] (222.93, 95.53) --
	(225.57, 90.96) --
	(220.29, 90.96) --
	cycle;

\path[draw=drawColor,line width= 0.4pt,line join=round,line cap=round] (264.79, 99.82) rectangle (268.71,103.74);

\path[draw=drawColor,line width= 0.4pt,line join=round,line cap=round] (264.79, 99.82) -- (268.71,103.74);

\path[draw=drawColor,line width= 0.4pt,line join=round,line cap=round] (264.79,103.74) -- (268.71, 99.82);

\path[draw=drawColor,line width= 0.4pt,line join=round,line cap=round] (277.15,109.11) rectangle (281.07,113.04);

\path[draw=drawColor,line width= 0.4pt,line join=round,line cap=round] (277.15,109.11) -- (281.07,113.04);

\path[draw=drawColor,line width= 0.4pt,line join=round,line cap=round] (277.15,113.04) -- (281.07,109.11);
\definecolor{drawColor}{RGB}{0,176,246}

\path[draw=drawColor,line width= 0.4pt,line join=round,line cap=round] (299.87,120.37) -- (305.41,120.37);

\path[draw=drawColor,line width= 0.4pt,line join=round,line cap=round] (302.64,117.59) -- (302.64,123.14);

\path[draw=drawColor,line width= 0.4pt,line join=round,line cap=round] (310.65,129.67) -- (316.20,129.67);

\path[draw=drawColor,line width= 0.4pt,line join=round,line cap=round] (313.42,126.89) -- (313.42,132.44);

\path[draw=drawColor,line width= 0.4pt,line join=round,line cap=round] (320.79,138.96) -- (326.34,138.96);

\path[draw=drawColor,line width= 0.4pt,line join=round,line cap=round] (323.56,136.19) -- (323.56,141.74);
\definecolor{drawColor}{RGB}{0,191,125}

\path[draw=drawColor,line width= 0.6pt,line join=round] (184.55, 26.94) --
	(184.55, 27.87);

\path[draw=drawColor,line width= 0.6pt,line join=round] (184.55, 27.41) --
	(175.78, 27.41);

\path[draw=drawColor,line width= 0.6pt,line join=round] (175.78, 26.94) --
	(175.78, 27.87);
\definecolor{drawColor}{RGB}{248,118,109}

\path[draw=drawColor,line width= 0.6pt,line join=round] (187.79, 36.24) --
	(187.79, 37.17);

\path[draw=drawColor,line width= 0.6pt,line join=round] (187.79, 36.70) --
	(179.45, 36.70);

\path[draw=drawColor,line width= 0.6pt,line join=round] (179.45, 36.24) --
	(179.45, 37.17);
\definecolor{drawColor}{RGB}{0,191,125}

\path[draw=drawColor,line width= 0.6pt,line join=round] (189.08, 45.54) --
	(189.08, 46.46);

\path[draw=drawColor,line width= 0.6pt,line join=round] (189.08, 46.00) --
	(179.71, 46.00);

\path[draw=drawColor,line width= 0.6pt,line join=round] (179.71, 45.54) --
	(179.71, 46.46);
\definecolor{drawColor}{RGB}{163,165,0}

\path[draw=drawColor,line width= 0.6pt,line join=round] (199.45, 54.83) --
	(199.45, 55.76);

\path[draw=drawColor,line width= 0.6pt,line join=round] (199.45, 55.30) --
	(189.62, 55.30);

\path[draw=drawColor,line width= 0.6pt,line join=round] (189.62, 54.83) --
	(189.62, 55.76);

\path[draw=drawColor,line width= 0.6pt,line join=round] (202.35, 64.13) --
	(202.35, 65.06);

\path[draw=drawColor,line width= 0.6pt,line join=round] (202.35, 64.59) --
	(192.20, 64.59);

\path[draw=drawColor,line width= 0.6pt,line join=round] (192.20, 64.13) --
	(192.20, 65.06);
\definecolor{drawColor}{RGB}{231,107,243}

\path[draw=drawColor,line width= 0.6pt,line join=round] (212.16, 73.42) --
	(212.16, 74.35);

\path[draw=drawColor,line width= 0.6pt,line join=round] (212.16, 73.89) --
	(201.11, 73.89);

\path[draw=drawColor,line width= 0.6pt,line join=round] (201.11, 73.42) --
	(201.11, 74.35);
\definecolor{drawColor}{RGB}{248,118,109}

\path[draw=drawColor,line width= 0.6pt,line join=round] (214.08, 82.72) --
	(214.08, 83.65);

\path[draw=drawColor,line width= 0.6pt,line join=round] (214.08, 83.18) --
	(204.65, 83.18);

\path[draw=drawColor,line width= 0.6pt,line join=round] (204.65, 82.72) --
	(204.65, 83.65);
\definecolor{drawColor}{RGB}{163,165,0}

\path[draw=drawColor,line width= 0.6pt,line join=round] (229.23, 92.02) --
	(229.23, 92.95);

\path[draw=drawColor,line width= 0.6pt,line join=round] (229.23, 92.48) --
	(216.63, 92.48);

\path[draw=drawColor,line width= 0.6pt,line join=round] (216.63, 92.02) --
	(216.63, 92.95);
\definecolor{drawColor}{RGB}{231,107,243}

\path[draw=drawColor,line width= 0.6pt,line join=round] (271.53,101.31) --
	(271.53,102.24);

\path[draw=drawColor,line width= 0.6pt,line join=round] (271.53,101.78) --
	(261.98,101.78);

\path[draw=drawColor,line width= 0.6pt,line join=round] (261.98,101.31) --
	(261.98,102.24);

\path[draw=drawColor,line width= 0.6pt,line join=round] (284.11,110.61) --
	(284.11,111.54);

\path[draw=drawColor,line width= 0.6pt,line join=round] (284.11,111.07) --
	(274.12,111.07);

\path[draw=drawColor,line width= 0.6pt,line join=round] (274.12,110.61) --
	(274.12,111.54);
\definecolor{drawColor}{RGB}{0,176,246}

\path[draw=drawColor,line width= 0.6pt,line join=round] (306.73,119.91) --
	(306.73,120.83);

\path[draw=drawColor,line width= 0.6pt,line join=round] (306.73,120.37) --
	(298.55,120.37);

\path[draw=drawColor,line width= 0.6pt,line join=round] (298.55,119.91) --
	(298.55,120.83);

\path[draw=drawColor,line width= 0.6pt,line join=round] (317.14,129.20) --
	(317.14,130.13);

\path[draw=drawColor,line width= 0.6pt,line join=round] (317.14,129.67) --
	(309.70,129.67);

\path[draw=drawColor,line width= 0.6pt,line join=round] (309.70,129.20) --
	(309.70,130.13);

\path[draw=drawColor,line width= 0.6pt,line join=round] (326.72,138.50) --
	(326.72,139.43);

\path[draw=drawColor,line width= 0.6pt,line join=round] (326.72,138.96) --
	(320.40,138.96);

\path[draw=drawColor,line width= 0.6pt,line join=round] (320.40,138.50) --
	(320.40,139.43);
\end{scope}
\begin{scope}
\path[clip] (  0.00,  0.00) rectangle (523.96,144.54);
\definecolor{drawColor}{RGB}{0,0,0}

\node[text=drawColor,anchor=base east,inner sep=0pt, outer sep=0pt, scale=  0.80] at (126.58, 24.65) {Political Stance / Ideology};

\node[text=drawColor,anchor=base east,inner sep=0pt, outer sep=0pt, scale=  0.80] at (126.58, 33.95) {Social Impact / Investing in Community};

\node[text=drawColor,anchor=base east,inner sep=0pt, outer sep=0pt, scale=  0.80] at (126.58, 43.25) {Diversity / Equality / Inclusivity};

\node[text=drawColor,anchor=base east,inner sep=0pt, outer sep=0pt, scale=  0.80] at (126.58, 52.54) {Country of Origin / Place of Manufacture};

\node[text=drawColor,anchor=base east,inner sep=0pt, outer sep=0pt, scale=  0.80] at (126.58, 61.84) {Ethical Sourcing / Production};

\node[text=drawColor,anchor=base east,inner sep=0pt, outer sep=0pt, scale=  0.80] at (126.58, 71.13) {Governance \& Compliance};

\node[text=drawColor,anchor=base east,inner sep=0pt, outer sep=0pt, scale=  0.80] at (126.58, 80.43) {Eco Friendly / Sustainable Practices};

\node[text=drawColor,anchor=base east,inner sep=0pt, outer sep=0pt, scale=  0.80] at (126.58, 89.73) {Slave/Child Labour / Employment Rights};

\node[text=drawColor,anchor=base east,inner sep=0pt, outer sep=0pt, scale=  0.80] at (126.58, 99.02) {Brand Appeal / Reputation};

\node[text=drawColor,anchor=base east,inner sep=0pt, outer sep=0pt, scale=  0.80] at (126.58,108.32) {Seller / Retailer};

\node[text=drawColor,anchor=base east,inner sep=0pt, outer sep=0pt, scale=  0.80] at (126.58,117.61) {Reviews / Opinions};

\node[text=drawColor,anchor=base east,inner sep=0pt, outer sep=0pt, scale=  0.80] at (126.58,126.91) {Price / Value};

\node[text=drawColor,anchor=base east,inner sep=0pt, outer sep=0pt, scale=  0.80] at (126.58,136.21) {Features / Quality};
\end{scope}
\begin{scope}
\path[clip] (  0.00,  0.00) rectangle (523.96,144.54);
\definecolor{drawColor}{gray}{0.20}

\path[draw=drawColor,line width= 0.6pt,line join=round] (128.78, 27.41) --
	(131.53, 27.41);

\path[draw=drawColor,line width= 0.6pt,line join=round] (128.78, 36.70) --
	(131.53, 36.70);

\path[draw=drawColor,line width= 0.6pt,line join=round] (128.78, 46.00) --
	(131.53, 46.00);

\path[draw=drawColor,line width= 0.6pt,line join=round] (128.78, 55.30) --
	(131.53, 55.30);

\path[draw=drawColor,line width= 0.6pt,line join=round] (128.78, 64.59) --
	(131.53, 64.59);

\path[draw=drawColor,line width= 0.6pt,line join=round] (128.78, 73.89) --
	(131.53, 73.89);

\path[draw=drawColor,line width= 0.6pt,line join=round] (128.78, 83.18) --
	(131.53, 83.18);

\path[draw=drawColor,line width= 0.6pt,line join=round] (128.78, 92.48) --
	(131.53, 92.48);

\path[draw=drawColor,line width= 0.6pt,line join=round] (128.78,101.78) --
	(131.53,101.78);

\path[draw=drawColor,line width= 0.6pt,line join=round] (128.78,111.07) --
	(131.53,111.07);

\path[draw=drawColor,line width= 0.6pt,line join=round] (128.78,120.37) --
	(131.53,120.37);

\path[draw=drawColor,line width= 0.6pt,line join=round] (128.78,129.67) --
	(131.53,129.67);

\path[draw=drawColor,line width= 0.6pt,line join=round] (128.78,138.96) --
	(131.53,138.96);
\end{scope}
\begin{scope}
\path[clip] (  0.00,  0.00) rectangle (523.96,144.54);
\definecolor{drawColor}{gray}{0.20}

\path[draw=drawColor,line width= 0.6pt,line join=round] (142.03, 19.08) --
	(142.03, 21.83);

\path[draw=drawColor,line width= 0.6pt,line join=round] (194.53, 19.08) --
	(194.53, 21.83);

\path[draw=drawColor,line width= 0.6pt,line join=round] (247.04, 19.08) --
	(247.04, 21.83);

\path[draw=drawColor,line width= 0.6pt,line join=round] (299.54, 19.08) --
	(299.54, 21.83);

\path[draw=drawColor,line width= 0.6pt,line join=round] (352.04, 19.08) --
	(352.04, 21.83);
\end{scope}
\begin{scope}
\path[clip] (  0.00,  0.00) rectangle (523.96,144.54);
\definecolor{drawColor}{RGB}{0,0,0}

\node[text=drawColor,anchor=base,inner sep=0pt, outer sep=0pt, scale=  0.80] at (142.03, 11.37) {1};

\node[text=drawColor,anchor=base,inner sep=0pt, outer sep=0pt, scale=  0.80] at (194.53, 11.37) {2};

\node[text=drawColor,anchor=base,inner sep=0pt, outer sep=0pt, scale=  0.80] at (247.04, 11.37) {3};

\node[text=drawColor,anchor=base,inner sep=0pt, outer sep=0pt, scale=  0.80] at (299.54, 11.37) {4};

\node[text=drawColor,anchor=base,inner sep=0pt, outer sep=0pt, scale=  0.80] at (352.04, 11.37) {5};
\end{scope}
\begin{scope}
\path[clip] (  0.00,  0.00) rectangle (523.96,144.54);
\definecolor{drawColor}{RGB}{0,0,0}

\node[text=drawColor,anchor=base,inner sep=0pt, outer sep=0pt, scale=  0.80] at (247.04,  1.56) {Importance};
\end{scope}
\begin{scope}
\path[clip] (  0.00,  0.00) rectangle (523.96,144.54);

\path[] (373.54, 38.02) rectangle (545.30,128.35);
\end{scope}
\begin{scope}
\path[clip] (  0.00,  0.00) rectangle (523.96,144.54);
\definecolor{drawColor}{RGB}{0,0,0}

\node[text=drawColor,anchor=base west,inner sep=0pt, outer sep=0pt, scale=  0.80] at (379.04,116.57) {Theme:};
\end{scope}
\begin{scope}
\path[clip] (  0.00,  0.00) rectangle (523.96,144.54);
\definecolor{fillColor}{gray}{0.95}

\path[fill=fillColor] (379.04,101.33) rectangle (393.50,115.79);
\end{scope}
\begin{scope}
\path[clip] (  0.00,  0.00) rectangle (523.96,144.54);
\definecolor{fillColor}{RGB}{248,118,109}

\path[fill=fillColor] (386.27,108.56) circle (  1.96);
\end{scope}
\begin{scope}
\path[clip] (  0.00,  0.00) rectangle (523.96,144.54);
\definecolor{drawColor}{RGB}{248,118,109}

\path[draw=drawColor,line width= 0.6pt,line join=round] (380.49,108.56) -- (392.05,108.56);
\end{scope}
\begin{scope}
\path[clip] (  0.00,  0.00) rectangle (523.96,144.54);
\definecolor{fillColor}{gray}{0.95}

\path[fill=fillColor] (379.04, 86.88) rectangle (393.50,101.33);
\end{scope}
\begin{scope}
\path[clip] (  0.00,  0.00) rectangle (523.96,144.54);
\definecolor{fillColor}{RGB}{163,165,0}

\path[fill=fillColor] (386.27, 97.16) --
	(388.91, 92.58) --
	(383.63, 92.58) --
	cycle;
\end{scope}
\begin{scope}
\path[clip] (  0.00,  0.00) rectangle (523.96,144.54);
\definecolor{drawColor}{RGB}{163,165,0}

\path[draw=drawColor,line width= 0.6pt,line join=round] (380.49, 94.11) -- (392.05, 94.11);
\end{scope}
\begin{scope}
\path[clip] (  0.00,  0.00) rectangle (523.96,144.54);
\definecolor{fillColor}{gray}{0.95}

\path[fill=fillColor] (379.04, 72.43) rectangle (393.50, 86.88);
\end{scope}
\begin{scope}
\path[clip] (  0.00,  0.00) rectangle (523.96,144.54);
\definecolor{fillColor}{RGB}{0,191,125}

\path[fill=fillColor] (384.31, 77.69) --
	(388.23, 77.69) --
	(388.23, 81.61) --
	(384.31, 81.61) --
	cycle;
\end{scope}
\begin{scope}
\path[clip] (  0.00,  0.00) rectangle (523.96,144.54);
\definecolor{drawColor}{RGB}{0,191,125}

\path[draw=drawColor,line width= 0.6pt,line join=round] (380.49, 79.65) -- (392.05, 79.65);
\end{scope}
\begin{scope}
\path[clip] (  0.00,  0.00) rectangle (523.96,144.54);
\definecolor{fillColor}{gray}{0.95}

\path[fill=fillColor] (379.04, 57.97) rectangle (393.50, 72.43);
\end{scope}
\begin{scope}
\path[clip] (  0.00,  0.00) rectangle (523.96,144.54);
\definecolor{drawColor}{RGB}{0,176,246}

\path[draw=drawColor,line width= 0.4pt,line join=round,line cap=round] (383.50, 65.20) -- (389.05, 65.20);

\path[draw=drawColor,line width= 0.4pt,line join=round,line cap=round] (386.27, 62.42) -- (386.27, 67.97);
\end{scope}
\begin{scope}
\path[clip] (  0.00,  0.00) rectangle (523.96,144.54);
\definecolor{drawColor}{RGB}{0,176,246}

\path[draw=drawColor,line width= 0.6pt,line join=round] (380.49, 65.20) -- (392.05, 65.20);
\end{scope}
\begin{scope}
\path[clip] (  0.00,  0.00) rectangle (523.96,144.54);
\definecolor{fillColor}{gray}{0.95}

\path[fill=fillColor] (379.04, 43.52) rectangle (393.50, 57.97);
\end{scope}
\begin{scope}
\path[clip] (  0.00,  0.00) rectangle (523.96,144.54);
\definecolor{drawColor}{RGB}{231,107,243}

\path[draw=drawColor,line width= 0.4pt,line join=round,line cap=round] (384.31, 48.78) rectangle (388.23, 52.71);

\path[draw=drawColor,line width= 0.4pt,line join=round,line cap=round] (384.31, 48.78) -- (388.23, 52.71);

\path[draw=drawColor,line width= 0.4pt,line join=round,line cap=round] (384.31, 52.71) -- (388.23, 48.78);
\end{scope}
\begin{scope}
\path[clip] (  0.00,  0.00) rectangle (523.96,144.54);
\definecolor{drawColor}{RGB}{231,107,243}

\path[draw=drawColor,line width= 0.6pt,line join=round] (380.49, 50.74) -- (392.05, 50.74);
\end{scope}
\begin{scope}
\path[clip] (  0.00,  0.00) rectangle (523.96,144.54);
\definecolor{drawColor}{RGB}{0,0,0}

\node[text=drawColor,anchor=base west,inner sep=0pt, outer sep=0pt, scale=  0.80] at (397.50,105.81) {Environmental and Social Responsibility};
\end{scope}
\begin{scope}
\path[clip] (  0.00,  0.00) rectangle (523.96,144.54);
\definecolor{drawColor}{RGB}{0,0,0}

\node[text=drawColor,anchor=base west,inner sep=0pt, outer sep=0pt, scale=  0.80] at (397.50, 91.35) {Ethical Procurement};
\end{scope}
\begin{scope}
\path[clip] (  0.00,  0.00) rectangle (523.96,144.54);
\definecolor{drawColor}{RGB}{0,0,0}

\node[text=drawColor,anchor=base west,inner sep=0pt, outer sep=0pt, scale=  0.80] at (397.50, 76.90) {Inclusion and Ideology};
\end{scope}
\begin{scope}
\path[clip] (  0.00,  0.00) rectangle (523.96,144.54);
\definecolor{drawColor}{RGB}{0,0,0}

\node[text=drawColor,anchor=base west,inner sep=0pt, outer sep=0pt, scale=  0.80] at (397.50, 62.44) {Product Evaluation};
\end{scope}
\begin{scope}
\path[clip] (  0.00,  0.00) rectangle (523.96,144.54);
\definecolor{drawColor}{RGB}{0,0,0}

\node[text=drawColor,anchor=base west,inner sep=0pt, outer sep=0pt, scale=  0.80] at (397.50, 47.99) {Reputation and Governance};
\end{scope}
\end{tikzpicture}

%% file: objects/searches_chart.tex
\begin{tikzpicture}[x=1pt,y=1pt]
\definecolor{fillColor}{RGB}{255,255,255}
\path[use as bounding box,fill=fillColor,fill opacity=0.00] (0,0) rectangle (505.89, 86.72);
\begin{scope}
\path[clip] (  0.00,  0.00) rectangle (505.89, 86.72);

\path[] (  0.00,  0.00) rectangle (505.89, 86.72);
\end{scope}
\begin{scope}
\path[clip] (125.91, 21.83) rectangle (347.06, 86.72);

\path[] (125.91, 21.83) rectangle (347.06, 86.72);
\definecolor{drawColor}{RGB}{211,211,211}

\path[draw=drawColor,line width= 0.6pt,line join=round] (125.91, 29.32) --
	(347.06, 29.32);

\path[draw=drawColor,line width= 0.6pt,line join=round] (125.91, 41.80) --
	(347.06, 41.80);

\path[draw=drawColor,line width= 0.6pt,line join=round] (125.91, 54.28) --
	(347.06, 54.28);

\path[draw=drawColor,line width= 0.6pt,line join=round] (125.91, 66.76) --
	(347.06, 66.76);

\path[draw=drawColor,line width= 0.6pt,line join=round] (125.91, 79.24) --
	(347.06, 79.24);

\path[draw=drawColor,line width= 0.6pt,line join=round] (135.96, 21.83) --
	(135.96, 86.72);

\path[draw=drawColor,line width= 0.6pt,line join=round] (186.22, 21.83) --
	(186.22, 86.72);

\path[draw=drawColor,line width= 0.6pt,line join=round] (236.49, 21.83) --
	(236.49, 86.72);

\path[draw=drawColor,line width= 0.6pt,line join=round] (286.75, 21.83) --
	(286.75, 86.72);

\path[draw=drawColor,line width= 0.6pt,line join=round] (337.01, 21.83) --
	(337.01, 86.72);
\definecolor{fillColor}{RGB}{0,191,196}

\path[fill=fillColor] (137.82, 36.18) rectangle (286.84, 47.41);
\definecolor{fillColor}{RGB}{199,124,255}

\path[fill=fillColor] (135.96, 36.18) rectangle (137.82, 47.41);
\definecolor{fillColor}{RGB}{248,118,109}

\path[fill=fillColor] (314.34, 36.18) rectangle (337.01, 47.41);
\definecolor{fillColor}{RGB}{124,174,0}

\path[fill=fillColor] (286.84, 36.18) rectangle (314.34, 47.41);
\definecolor{fillColor}{RGB}{0,191,196}

\path[fill=fillColor] (137.13, 48.66) rectangle (284.52, 59.89);
\definecolor{fillColor}{RGB}{199,124,255}

\path[fill=fillColor] (135.96, 48.66) rectangle (137.13, 59.89);
\definecolor{fillColor}{RGB}{248,118,109}

\path[fill=fillColor] (314.28, 48.66) rectangle (337.01, 59.89);
\definecolor{fillColor}{RGB}{124,174,0}

\path[fill=fillColor] (284.52, 48.66) rectangle (314.28, 59.89);
\definecolor{fillColor}{RGB}{0,191,196}

\path[fill=fillColor] (137.09, 23.70) rectangle (305.94, 34.93);
\definecolor{fillColor}{RGB}{199,124,255}

\path[fill=fillColor] (135.96, 23.70) rectangle (137.09, 34.93);
\definecolor{fillColor}{RGB}{248,118,109}

\path[fill=fillColor] (323.53, 23.70) rectangle (337.01, 34.93);
\definecolor{fillColor}{RGB}{124,174,0}

\path[fill=fillColor] (305.94, 23.70) rectangle (323.53, 34.93);
\definecolor{fillColor}{RGB}{0,191,196}

\path[fill=fillColor] (136.21, 73.62) rectangle (142.97, 84.85);
\definecolor{fillColor}{RGB}{199,124,255}

\path[fill=fillColor] (135.96, 73.62) rectangle (136.21, 84.85);
\definecolor{fillColor}{RGB}{248,118,109}

\path[fill=fillColor] (314.23, 73.62) rectangle (337.01, 84.85);
\definecolor{fillColor}{RGB}{124,174,0}

\path[fill=fillColor] (142.97, 73.62) rectangle (314.23, 84.85);
\definecolor{fillColor}{RGB}{0,191,196}

\path[fill=fillColor] (137.19, 61.14) rectangle (201.01, 72.37);
\definecolor{fillColor}{RGB}{199,124,255}

\path[fill=fillColor] (135.96, 61.14) rectangle (137.19, 72.37);
\definecolor{fillColor}{RGB}{248,118,109}

\path[fill=fillColor] (292.09, 61.14) rectangle (337.01, 72.37);
\definecolor{fillColor}{RGB}{124,174,0}

\path[fill=fillColor] (201.01, 61.14) rectangle (292.09, 72.37);
\end{scope}
\begin{scope}
\path[clip] (  0.00,  0.00) rectangle (505.89, 86.72);
\definecolor{drawColor}{RGB}{0,0,0}

\node[text=drawColor,anchor=base east,inner sep=0pt, outer sep=0pt, scale=  0.80] at (120.96, 26.56) {Inclusion and Ideology};

\node[text=drawColor,anchor=base east,inner sep=0pt, outer sep=0pt, scale=  0.80] at (120.96, 39.04) {Environmental and Social Responsibility};

\node[text=drawColor,anchor=base east,inner sep=0pt, outer sep=0pt, scale=  0.80] at (120.96, 51.52) {Ethical Procurement};

\node[text=drawColor,anchor=base east,inner sep=0pt, outer sep=0pt, scale=  0.80] at (120.96, 64.00) {Reputation and Governance};

\node[text=drawColor,anchor=base east,inner sep=0pt, outer sep=0pt, scale=  0.80] at (120.96, 76.48) {Product Evaluation};
\end{scope}
\begin{scope}
\path[clip] (  0.00,  0.00) rectangle (505.89, 86.72);
\definecolor{drawColor}{gray}{0.20}

\path[draw=drawColor,line width= 0.6pt,line join=round] (123.16, 29.32) --
	(125.91, 29.32);

\path[draw=drawColor,line width= 0.6pt,line join=round] (123.16, 41.80) --
	(125.91, 41.80);

\path[draw=drawColor,line width= 0.6pt,line join=round] (123.16, 54.28) --
	(125.91, 54.28);

\path[draw=drawColor,line width= 0.6pt,line join=round] (123.16, 66.76) --
	(125.91, 66.76);

\path[draw=drawColor,line width= 0.6pt,line join=round] (123.16, 79.24) --
	(125.91, 79.24);
\end{scope}
\begin{scope}
\path[clip] (  0.00,  0.00) rectangle (505.89, 86.72);
\definecolor{drawColor}{gray}{0.20}

\path[draw=drawColor,line width= 0.6pt,line join=round] (135.96, 19.08) --
	(135.96, 21.83);

\path[draw=drawColor,line width= 0.6pt,line join=round] (186.22, 19.08) --
	(186.22, 21.83);

\path[draw=drawColor,line width= 0.6pt,line join=round] (236.49, 19.08) --
	(236.49, 21.83);

\path[draw=drawColor,line width= 0.6pt,line join=round] (286.75, 19.08) --
	(286.75, 21.83);

\path[draw=drawColor,line width= 0.6pt,line join=round] (337.01, 19.08) --
	(337.01, 21.83);
\end{scope}
\begin{scope}
\path[clip] (  0.00,  0.00) rectangle (505.89, 86.72);
\definecolor{drawColor}{RGB}{0,0,0}

\node[text=drawColor,anchor=base,inner sep=0pt, outer sep=0pt, scale=  0.80] at (135.96, 11.37) {0\%};

\node[text=drawColor,anchor=base,inner sep=0pt, outer sep=0pt, scale=  0.80] at (186.22, 11.37) {25\%};

\node[text=drawColor,anchor=base,inner sep=0pt, outer sep=0pt, scale=  0.80] at (236.49, 11.37) {50\%};

\node[text=drawColor,anchor=base,inner sep=0pt, outer sep=0pt, scale=  0.80] at (286.75, 11.37) {75\%};

\node[text=drawColor,anchor=base,inner sep=0pt, outer sep=0pt, scale=  0.80] at (337.01, 11.37) {100\%};
\end{scope}
\begin{scope}
\path[clip] (  0.00,  0.00) rectangle (505.89, 86.72);
\definecolor{drawColor}{RGB}{0,0,0}

\node[text=drawColor,anchor=base,inner sep=0pt, outer sep=0pt, scale=  0.80] at (236.49,  1.56) {Percentage};
\end{scope}
\begin{scope}
\path[clip] (  0.00,  0.00) rectangle (505.89, 86.72);

\path[] (358.06, 19.15) rectangle (484.55, 89.40);
\end{scope}
\begin{scope}
\path[clip] (  0.00,  0.00) rectangle (505.89, 86.72);
\definecolor{drawColor}{RGB}{0,0,0}

\node[text=drawColor,anchor=base west,inner sep=0pt, outer sep=0pt, scale=  0.80] at (363.56, 77.61) {Activity:};
\end{scope}
\begin{scope}
\path[clip] (  0.00,  0.00) rectangle (505.89, 86.72);
\definecolor{fillColor}{gray}{0.95}

\path[fill=fillColor] (363.56, 60.79) rectangle (375.61, 72.83);
\end{scope}
\begin{scope}
\path[clip] (  0.00,  0.00) rectangle (505.89, 86.72);
\definecolor{fillColor}{RGB}{248,118,109}

\path[fill=fillColor] (364.27, 61.50) rectangle (374.90, 72.12);
\end{scope}
\begin{scope}
\path[clip] (  0.00,  0.00) rectangle (505.89, 86.72);
\definecolor{fillColor}{gray}{0.95}

\path[fill=fillColor] (363.56, 48.74) rectangle (375.61, 60.79);
\end{scope}
\begin{scope}
\path[clip] (  0.00,  0.00) rectangle (505.89, 86.72);
\definecolor{fillColor}{RGB}{124,174,0}

\path[fill=fillColor] (364.27, 49.46) rectangle (374.90, 60.08);
\end{scope}
\begin{scope}
\path[clip] (  0.00,  0.00) rectangle (505.89, 86.72);
\definecolor{fillColor}{gray}{0.95}

\path[fill=fillColor] (363.56, 36.70) rectangle (375.61, 48.74);
\end{scope}
\begin{scope}
\path[clip] (  0.00,  0.00) rectangle (505.89, 86.72);
\definecolor{fillColor}{RGB}{0,191,196}

\path[fill=fillColor] (364.27, 37.41) rectangle (374.90, 48.03);
\end{scope}
\begin{scope}
\path[clip] (  0.00,  0.00) rectangle (505.89, 86.72);
\definecolor{fillColor}{gray}{0.95}

\path[fill=fillColor] (363.56, 24.65) rectangle (375.61, 36.70);
\end{scope}
\begin{scope}
\path[clip] (  0.00,  0.00) rectangle (505.89, 86.72);
\definecolor{fillColor}{RGB}{199,124,255}

\path[fill=fillColor] (364.27, 25.37) rectangle (374.90, 35.99);
\end{scope}
\begin{scope}
\path[clip] (  0.00,  0.00) rectangle (505.89, 86.72);
\definecolor{drawColor}{RGB}{0,0,0}

\node[text=drawColor,anchor=base west,inner sep=0pt, outer sep=0pt, scale=  0.80] at (379.61, 64.06) {considered, not searched};
\end{scope}
\begin{scope}
\path[clip] (  0.00,  0.00) rectangle (505.89, 86.72);
\definecolor{drawColor}{RGB}{0,0,0}

\node[text=drawColor,anchor=base west,inner sep=0pt, outer sep=0pt, scale=  0.80] at (379.61, 52.01) {considered, searched};
\end{scope}
\begin{scope}
\path[clip] (  0.00,  0.00) rectangle (505.89, 86.72);
\definecolor{drawColor}{RGB}{0,0,0}

\node[text=drawColor,anchor=base west,inner sep=0pt, outer sep=0pt, scale=  0.80] at (379.61, 39.97) {not considered, not searched};
\end{scope}
\begin{scope}
\path[clip] (  0.00,  0.00) rectangle (505.89, 86.72);
\definecolor{drawColor}{RGB}{0,0,0}

\node[text=drawColor,anchor=base west,inner sep=0pt, outer sep=0pt, scale=  0.80] at (379.61, 27.92) {not considered, searched};
\end{scope}
\end{tikzpicture}

%% file: objects/brehm_chart.tex
\begin{tikzpicture}[x=1pt,y=1pt]
\definecolor{fillColor}{RGB}{255,255,255}
\path[use as bounding box,fill=fillColor,fill opacity=0.00] (0,0) rectangle (505.89,126.47);
\begin{scope}
\path[clip] (  0.00,  0.00) rectangle (505.89,126.47);

\path[] (  0.00,  0.00) rectangle (505.89,126.47);
\end{scope}
\begin{scope}
\path[clip] (125.91, 21.83) rectangle (370.09,126.47);

\path[] (125.91, 21.83) rectangle (370.09,126.47);
\definecolor{drawColor}{RGB}{211,211,211}

\path[draw=drawColor,line width= 0.6pt,line join=round] (125.91, 33.90) --
	(370.09, 33.90);

\path[draw=drawColor,line width= 0.6pt,line join=round] (125.91, 54.03) --
	(370.09, 54.03);

\path[draw=drawColor,line width= 0.6pt,line join=round] (125.91, 74.15) --
	(370.09, 74.15);

\path[draw=drawColor,line width= 0.6pt,line join=round] (125.91, 94.27) --
	(370.09, 94.27);

\path[draw=drawColor,line width= 0.6pt,line join=round] (125.91,114.40) --
	(370.09,114.40);

\path[draw=drawColor,line width= 0.6pt,line join=round] (183.51, 21.83) --
	(183.51,126.47);

\path[draw=drawColor,line width= 0.6pt,line join=round] (251.48, 21.83) --
	(251.48,126.47);

\path[draw=drawColor,line width= 0.6pt,line join=round] (319.45, 21.83) --
	(319.45,126.47);
\definecolor{fillColor}{RGB}{248,118,109}

\path[fill=fillColor] (158.85, 68.92) circle (  1.96);

\path[fill=fillColor] (196.73, 48.80) circle (  1.96);

\path[fill=fillColor] (148.42, 28.67) circle (  1.96);

\path[fill=fillColor] (345.91,109.17) circle (  1.96);

\path[fill=fillColor] (286.52, 89.04) circle (  1.96);
\definecolor{fillColor}{RGB}{97,156,255}

\path[fill=fillColor] (162.62, 77.42) --
	(166.55, 77.42) --
	(166.55, 81.35) --
	(162.62, 81.35) --
	cycle;

\path[fill=fillColor] (193.05, 57.30) --
	(196.98, 57.30) --
	(196.98, 61.22) --
	(193.05, 61.22) --
	cycle;

\path[fill=fillColor] (176.44, 37.17) --
	(180.37, 37.17) --
	(180.37, 41.10) --
	(176.44, 41.10) --
	cycle;

\path[fill=fillColor] (350.48,117.67) --
	(354.40,117.67) --
	(354.40,121.59) --
	(350.48,121.59) --
	cycle;

\path[fill=fillColor] (254.80, 97.54) --
	(258.73, 97.54) --
	(258.73,101.47) --
	(254.80,101.47) --
	cycle;
\definecolor{fillColor}{RGB}{0,186,56}

\path[fill=fillColor] (167.45, 77.20) --
	(170.10, 72.63) --
	(164.81, 72.63) --
	cycle;

\path[fill=fillColor] (201.88, 57.08) --
	(204.52, 52.50) --
	(199.24, 52.50) --
	cycle;

\path[fill=fillColor] (169.52, 36.96) --
	(172.16, 32.38) --
	(166.88, 32.38) --
	cycle;

\path[fill=fillColor] (329.78,117.45) --
	(332.42,112.87) --
	(327.14,112.87) --
	cycle;

\path[fill=fillColor] (275.65, 97.33) --
	(278.29, 92.75) --
	(273.00, 92.75) --
	cycle;
\definecolor{drawColor}{RGB}{248,118,109}

\path[draw=drawColor,line width= 0.6pt,line join=round] (170.72, 68.58) --
	(170.72, 69.25);

\path[draw=drawColor,line width= 0.6pt,line join=round] (170.72, 68.92) --
	(146.98, 68.92);

\path[draw=drawColor,line width= 0.6pt,line join=round] (146.98, 68.58) --
	(146.98, 69.25);

\path[draw=drawColor,line width= 0.6pt,line join=round] (206.82, 48.46) --
	(206.82, 49.13);

\path[draw=drawColor,line width= 0.6pt,line join=round] (206.82, 48.80) --
	(186.64, 48.80);

\path[draw=drawColor,line width= 0.6pt,line join=round] (186.64, 48.46) --
	(186.64, 49.13);

\path[draw=drawColor,line width= 0.6pt,line join=round] (159.83, 28.34) --
	(159.83, 29.01);

\path[draw=drawColor,line width= 0.6pt,line join=round] (159.83, 28.67) --
	(137.01, 28.67);

\path[draw=drawColor,line width= 0.6pt,line join=round] (137.01, 28.34) --
	(137.01, 29.01);

\path[draw=drawColor,line width= 0.6pt,line join=round] (353.19,108.83) --
	(353.19,109.50);

\path[draw=drawColor,line width= 0.6pt,line join=round] (353.19,109.17) --
	(338.62,109.17);

\path[draw=drawColor,line width= 0.6pt,line join=round] (338.62,108.83) --
	(338.62,109.50);

\path[draw=drawColor,line width= 0.6pt,line join=round] (294.49, 88.71) --
	(294.49, 89.38);

\path[draw=drawColor,line width= 0.6pt,line join=round] (294.49, 89.04) --
	(278.56, 89.04);

\path[draw=drawColor,line width= 0.6pt,line join=round] (278.56, 88.71) --
	(278.56, 89.38);
\definecolor{drawColor}{RGB}{97,156,255}

\path[draw=drawColor,line width= 0.6pt,line join=round] (173.98, 79.05) --
	(173.98, 79.72);

\path[draw=drawColor,line width= 0.6pt,line join=round] (173.98, 79.38) --
	(155.20, 79.38);

\path[draw=drawColor,line width= 0.6pt,line join=round] (155.20, 79.05) --
	(155.20, 79.72);

\path[draw=drawColor,line width= 0.6pt,line join=round] (204.36, 58.92) --
	(204.36, 59.60);

\path[draw=drawColor,line width= 0.6pt,line join=round] (204.36, 59.26) --
	(185.67, 59.26);

\path[draw=drawColor,line width= 0.6pt,line join=round] (185.67, 58.92) --
	(185.67, 59.60);

\path[draw=drawColor,line width= 0.6pt,line join=round] (186.95, 38.80) --
	(186.95, 39.47);

\path[draw=drawColor,line width= 0.6pt,line join=round] (186.95, 39.14) --
	(169.86, 39.14);

\path[draw=drawColor,line width= 0.6pt,line join=round] (169.86, 38.80) --
	(169.86, 39.47);

\path[draw=drawColor,line width= 0.6pt,line join=round] (358.99,119.30) --
	(358.99,119.97);

\path[draw=drawColor,line width= 0.6pt,line join=round] (358.99,119.63) --
	(345.88,119.63);

\path[draw=drawColor,line width= 0.6pt,line join=round] (345.88,119.30) --
	(345.88,119.97);

\path[draw=drawColor,line width= 0.6pt,line join=round] (264.31, 99.17) --
	(264.31, 99.84);

\path[draw=drawColor,line width= 0.6pt,line join=round] (264.31, 99.51) --
	(249.22, 99.51);

\path[draw=drawColor,line width= 0.6pt,line join=round] (249.22, 99.17) --
	(249.22, 99.84);
\definecolor{drawColor}{RGB}{0,186,56}

\path[draw=drawColor,line width= 0.6pt,line join=round] (178.94, 73.82) --
	(178.94, 74.49);

\path[draw=drawColor,line width= 0.6pt,line join=round] (178.94, 74.15) --
	(155.97, 74.15);

\path[draw=drawColor,line width= 0.6pt,line join=round] (155.97, 73.82) --
	(155.97, 74.49);

\path[draw=drawColor,line width= 0.6pt,line join=round] (211.71, 53.69) --
	(211.71, 54.36);

\path[draw=drawColor,line width= 0.6pt,line join=round] (211.71, 54.03) --
	(192.05, 54.03);

\path[draw=drawColor,line width= 0.6pt,line join=round] (192.05, 53.69) --
	(192.05, 54.36);

\path[draw=drawColor,line width= 0.6pt,line join=round] (180.46, 33.57) --
	(180.46, 34.24);

\path[draw=drawColor,line width= 0.6pt,line join=round] (180.46, 33.90) --
	(158.58, 33.90);

\path[draw=drawColor,line width= 0.6pt,line join=round] (158.58, 33.57) --
	(158.58, 34.24);

\path[draw=drawColor,line width= 0.6pt,line join=round] (337.19,114.06) --
	(337.19,114.73);

\path[draw=drawColor,line width= 0.6pt,line join=round] (337.19,114.40) --
	(322.38,114.40);

\path[draw=drawColor,line width= 0.6pt,line join=round] (322.38,114.06) --
	(322.38,114.73);

\path[draw=drawColor,line width= 0.6pt,line join=round] (283.82, 93.94) --
	(283.82, 94.61);

\path[draw=drawColor,line width= 0.6pt,line join=round] (283.82, 94.27) --
	(267.47, 94.27);

\path[draw=drawColor,line width= 0.6pt,line join=round] (267.47, 93.94) --
	(267.47, 94.61);
\end{scope}
\begin{scope}
\path[clip] (  0.00,  0.00) rectangle (505.89,126.47);
\definecolor{drawColor}{RGB}{0,0,0}

\node[text=drawColor,anchor=base east,inner sep=0pt, outer sep=0pt, scale=  0.80] at (120.96, 31.15) {Inclusion and Ideology};

\node[text=drawColor,anchor=base east,inner sep=0pt, outer sep=0pt, scale=  0.80] at (120.96, 51.27) {Ethical Procurement};

\node[text=drawColor,anchor=base east,inner sep=0pt, outer sep=0pt, scale=  0.80] at (120.96, 71.40) {Environmental and Social Responsibility};

\node[text=drawColor,anchor=base east,inner sep=0pt, outer sep=0pt, scale=  0.80] at (120.96, 91.52) {Reputation and Governance};

\node[text=drawColor,anchor=base east,inner sep=0pt, outer sep=0pt, scale=  0.80] at (120.96,111.64) {Product Evaluation};
\end{scope}
\begin{scope}
\path[clip] (  0.00,  0.00) rectangle (505.89,126.47);
\definecolor{drawColor}{gray}{0.20}

\path[draw=drawColor,line width= 0.6pt,line join=round] (123.16, 33.90) --
	(125.91, 33.90);

\path[draw=drawColor,line width= 0.6pt,line join=round] (123.16, 54.03) --
	(125.91, 54.03);

\path[draw=drawColor,line width= 0.6pt,line join=round] (123.16, 74.15) --
	(125.91, 74.15);

\path[draw=drawColor,line width= 0.6pt,line join=round] (123.16, 94.27) --
	(125.91, 94.27);

\path[draw=drawColor,line width= 0.6pt,line join=round] (123.16,114.40) --
	(125.91,114.40);
\end{scope}
\begin{scope}
\path[clip] (  0.00,  0.00) rectangle (505.89,126.47);
\definecolor{drawColor}{gray}{0.20}

\path[draw=drawColor,line width= 0.6pt,line join=round] (183.51, 19.08) --
	(183.51, 21.83);

\path[draw=drawColor,line width= 0.6pt,line join=round] (251.48, 19.08) --
	(251.48, 21.83);

\path[draw=drawColor,line width= 0.6pt,line join=round] (319.45, 19.08) --
	(319.45, 21.83);
\end{scope}
\begin{scope}
\path[clip] (  0.00,  0.00) rectangle (505.89,126.47);
\definecolor{drawColor}{RGB}{0,0,0}

\node[text=drawColor,anchor=base,inner sep=0pt, outer sep=0pt, scale=  0.80] at (183.51, 11.37) {5.2};

\node[text=drawColor,anchor=base,inner sep=0pt, outer sep=0pt, scale=  0.80] at (251.48, 11.37) {5.6};

\node[text=drawColor,anchor=base,inner sep=0pt, outer sep=0pt, scale=  0.80] at (319.45, 11.37) {6.0};
\end{scope}
\begin{scope}
\path[clip] (  0.00,  0.00) rectangle (505.89,126.47);
\definecolor{drawColor}{RGB}{0,0,0}

\node[text=drawColor,anchor=base,inner sep=0pt, outer sep=0pt, scale=  0.80] at (248.00,  1.56) {Likert scale response};
\end{scope}
\begin{scope}
\path[clip] (  0.00,  0.00) rectangle (505.89,126.47);

\path[] (381.09, 43.44) rectangle (484.55,104.86);
\end{scope}
\begin{scope}
\path[clip] (  0.00,  0.00) rectangle (505.89,126.47);
\definecolor{drawColor}{RGB}{0,0,0}

\node[text=drawColor,anchor=base west,inner sep=0pt, outer sep=0pt, scale=  0.80] at (386.59, 93.08) {Scale:};
\end{scope}
\begin{scope}
\path[clip] (  0.00,  0.00) rectangle (505.89,126.47);
\definecolor{fillColor}{gray}{0.95}

\path[fill=fillColor] (386.59, 77.85) rectangle (401.05, 92.30);
\end{scope}
\begin{scope}
\path[clip] (  0.00,  0.00) rectangle (505.89,126.47);
\definecolor{fillColor}{RGB}{248,118,109}

\path[fill=fillColor] (393.82, 85.07) circle (  1.96);
\end{scope}
\begin{scope}
\path[clip] (  0.00,  0.00) rectangle (505.89,126.47);
\definecolor{drawColor}{RGB}{248,118,109}

\path[draw=drawColor,line width= 0.6pt,line join=round] (388.04, 85.07) -- (399.60, 85.07);
\end{scope}
\begin{scope}
\path[clip] (  0.00,  0.00) rectangle (505.89,126.47);
\definecolor{fillColor}{gray}{0.95}

\path[fill=fillColor] (386.59, 63.39) rectangle (401.05, 77.85);
\end{scope}
\begin{scope}
\path[clip] (  0.00,  0.00) rectangle (505.89,126.47);
\definecolor{fillColor}{RGB}{0,186,56}

\path[fill=fillColor] (393.82, 73.67) --
	(396.46, 69.09) --
	(391.18, 69.09) --
	cycle;
\end{scope}
\begin{scope}
\path[clip] (  0.00,  0.00) rectangle (505.89,126.47);
\definecolor{drawColor}{RGB}{0,186,56}

\path[draw=drawColor,line width= 0.6pt,line join=round] (388.04, 70.62) -- (399.60, 70.62);
\end{scope}
\begin{scope}
\path[clip] (  0.00,  0.00) rectangle (505.89,126.47);
\definecolor{fillColor}{gray}{0.95}

\path[fill=fillColor] (386.59, 48.94) rectangle (401.05, 63.39);
\end{scope}
\begin{scope}
\path[clip] (  0.00,  0.00) rectangle (505.89,126.47);
\definecolor{fillColor}{RGB}{97,156,255}

\path[fill=fillColor] (391.86, 54.20) --
	(395.78, 54.20) --
	(395.78, 58.13) --
	(391.86, 58.13) --
	cycle;
\end{scope}
\begin{scope}
\path[clip] (  0.00,  0.00) rectangle (505.89,126.47);
\definecolor{drawColor}{RGB}{97,156,255}

\path[draw=drawColor,line width= 0.6pt,line join=round] (388.04, 56.16) -- (399.60, 56.16);
\end{scope}
\begin{scope}
\path[clip] (  0.00,  0.00) rectangle (505.89,126.47);
\definecolor{drawColor}{RGB}{0,0,0}

\node[text=drawColor,anchor=base west,inner sep=0pt, outer sep=0pt, scale=  0.80] at (405.05, 82.32) {Perceived Ease};
\end{scope}
\begin{scope}
\path[clip] (  0.00,  0.00) rectangle (505.89,126.47);
\definecolor{drawColor}{RGB}{0,0,0}

\node[text=drawColor,anchor=base west,inner sep=0pt, outer sep=0pt, scale=  0.80] at (405.05, 67.86) {Success Expectancies};
\end{scope}
\begin{scope}
\path[clip] (  0.00,  0.00) rectangle (505.89,126.47);
\definecolor{drawColor}{RGB}{0,0,0}

\node[text=drawColor,anchor=base west,inner sep=0pt, outer sep=0pt, scale=  0.80] at (405.05, 53.41) {Perceived Value};
\end{scope}
\end{tikzpicture}

%% file: 5_discussion.tex

\section{Discussion and Conclusion} \label{sec_discussion}
The primary objective of this study was to investigate the role of search in the context of socially responsible consumerism.
To achieve this goal, a survey was designed and administered involving 286 participants. Within this survey, participants were asked to provide insights into their decision-making processes concerning thirteen distinct aspects ranging from price to ideology. The approach adopted in this study diverges from previous research on socially responsible consumption, as it specifically focuses on the role of search within the broader purchase decision-making process. 

%
The outcomes of our investigation unveiled a notable pattern: 
participants, on the whole, did not give a strong priority to aspects associated with socially responsible consumption when shopping. This phenomenon was accompanied by the emergence of a progressively widening gap that extended from their initial intentions to subsequent stages, including importance valuation, considerations, and search behaviours pertaining to these specific aspects. This is despite the fact that according to the \EMCB scale most participants were considered to be strongly or somewhat ethically minded (in line with past studies ~\cite{Hasanzade2018SelectingShopping,Casais2022ThePriorities}).
Our findings suggests that when individuals are faced with actual purchase decisions, both their considerations of socially responsible aspects and their behaviour can be impeded by information and search-related challenges.
This implies that a major obstacle for consumers wanting to be socially responsible, at least in part, can be characterized as an information and search problem.
Our analysis identified four distinct challenges that partly contributed to this expanding gap.
%

%
%
Before delving into these challenges, several limitations should be considered when interpreting the results of this study. 
First, similar to other surveys related to socially responsible consumerism, we constructed a tailored survey utilizing existing questions and questionnaires wherever possible. As illustrated by the low alpha value for our effort scale, in some cases this approach has limited reliability. Concurrent validity was nonetheless attested by the expected correlation between scales.
Second, the validity of surveyed expectations and reasons for (not) searching may be susceptible to recall bias and recent experiences' influence. Participants provided retrospective evaluations that might not accurately reflect their immediate thought processes during actual purchases\footnote{\scriptsize{As an aside, when reviewing the comments by participant's we found them to be very open, honest and direct regarding their concern or lack there of.}}.  
Thirdly, although the study attempted to minimize social desirability bias\cite{Krumpal2013DeterminantsReview} by asking participants to comment on concrete recent purchase decisions rather than on hypothetical scenarios, some level of this bias may still be present in their responses.
Despite these limitations, the current research gave an initial exploration of the extent to which search moderates socially responsible consumption decisions. 
Future research may consider additional methods or experimental designs to extend this exploration and refine our understanding of responsible consumption behavior.

The extensiveness of purchase decisions and the highlighted role of search in these decisions suggest that they can be regarded as complex search problems.
%
Complex search problems typically exhibit extensive information needs, multiple criteria, and evolving user intents \cite{Wildemuth2014UntanglingStudies, Liu2020IdentifyingTasks,Choi2019TheComplexity,Capra2018}.
%
In our investigation, socially responsible decisions involve a multitude of information needs related to aspects including worker rights \& conditions, the sustainable and ecological sourcing \& production,  social impact \& investing in the community, environmental impact, and many more~\cite{Jones2022TheDifference}.
%
Participants demonstrated the challenge of reconciling various criteria, such as cost and product quality with the ethical aspects, which highlights the multifaceted nature of complex search tasks. Moreover, the evolving user intents we observed, particularly the progressive gap from intentions to actual search behaviour, are reminiscent of the dynamic nature of complex search tasks\cite{Liu2020IdentifyingTasks}. 
These parallels between our results and the characteristics of complex search tasks imply that the act of responsible consumption inherently involves intricate and extensive interactive information retrieval and search processes.
%
It suggests that complex search tasks in the context of socially responsible shopping can benefit from established solutions in information retrieval systems and interfaces~\cite{Hearst2006DesignInterfaces}.
For instance, faceted navigation techniques could help users increase their awareness of secondary aspects ~\cite{Hearst2006DesignInterfaces}, while information integration can provide a comprehensive representation of different alternatives, aiding users in making informed decisions~\citep{Choi2021OrgBox:Tasks,Crescenzi2021SupportingOrgBox}.
Nevertheless, as indicated by our participants, there is limited unbiased information available for some secondary aspects, which suggests a need for social-collaborative solutions as well \citep{Soulier2013ASearch,vanderSluis2022ARetrieval}.

The challenges described by participants in responsible consumption decisions reveal a fundamental issue related to knowledge calibration \citep{Alba2000KnowledgeKnow}.
Knowledge calibration is the process of aligning self-assessed knowledge with the validity of that knowledge.
An informed purchase decision presupposes a complete understanding of the aspects involved, whereas engaging in search necessitates a sense of meta-cognitive uncertainty about one's knowledge~\cite{Flavell1979MetacognitionInquiry.}.
Concerns about the completeness, accuracy, or relevance of one's knowledge each contribute to a sense of uncertainty conducive to search engagement \citep{Tormala2016ThePersuasion}.
However, studies indicate that individuals often exhibit overconfidence, particularly when faced with complex problems like socially responsible purchase decisions~\cite{Alba2000KnowledgeKnow}.
This overconfidence becomes, amongst others, apparent through omission neglect, where individuals remain unaware of aspects they fail to consider \citep{Pfeiffer2014EffectsEvaluation,Kardes2006DebiasingNeglect}, and in the mis-calibrated appreciation of the (ir)relevance of certain aspects ( \citep{Tormala2016ThePersuasion}).
Both of these cases were evident in our results, with participants indicating that they had forgotten to consider certain aspects or assuming that an aspect was irrelevant to the product category they were considering.
This optimism appears unfounded, considering the often lengthy and distributed supply chains associated with various products.
These issues suggest a potential deficiency in knowledge calibration.
To address this meta-cognitive uncertainty, various interventions can be employed, including prompts, questions, or simple reminders \citep{Kammerer2020TrainingsSearch} that collectively offer a form of knowledge context \citep{smith2019,Kammerer2020TrainingsSearch,Crescenzi2021SupportingOrgBox}.
These interventions likely become particularly relevant for participants who express moderate intentions for socially responsible consumption but still struggle to consider or search for specific aspects.
This discussion highlights the importance of assessing search interventions for their effectiveness in the accurate calibration of relevant decision-making aspects \citep[cf.][]{White2015BeliefSearch}.

Our findings highlight the role of easy access to relevant information in promoting socially responsible consumption. Participants' responses indicate that, within their decision-making processes, the perceived value of acquiring information related to socially responsible aspects often fell short of the associated costs. Here they faced challenges ranging from the unavailability and difficulty in finding relevant information to the insufficiency of their prior knowledge for effectively using available information. Moreover, participants express concerns about the reliability of the information they do encounter. These barriers collectively deter informed decision-making and undermine the incorporation of socially responsible aspects into their purchasing decisions. Ease of access and information reliability frequently posed challenges, not only limiting participants' ability to engage in searches but also appearing to lead to \textit{a priori} neglect of important aspects.

Addressing these challenges will be paramount in enabling consumers to be socially responsible. This will require fostering and building a compelling argument for the impact of these issues on society and the feasibility of addressing them. 
This work aims to serve as a foundation for change towards the creation of marketplaces where responsible decisions are both feasible and convenient.
Considering socially responsible shopping as an information and search problem partly shifts the onus of responsibility to information suppliers (e.g. producers of goods). Responsible (AI) based E-commerce platforms and search systems can then play an essential role in addressing these challenges by effectively communicating potential concerns, enhancing awareness among consumers, and offering product-domain explanations, and ensure the reliability of the information provided.
Acknowledging responsible consumption as an information and search problem should not overshadow its predominant economic underpinnings.
For many consumers, the complexities of socially responsible aspects may seem secondary to more immediate concerns, which means socially responsible consumption cannot solely be viewed as an information and search problem.  Both elements need to coalesce to shape the landscape of socially responsible consumerism. This will not only empowers consumers to make more informed decisions but will also foster and reward socially responsible production.
%
%
The journey towards more socially responsible is contingent on bridging the information gaps and fostering a more informed and conscious consumer base. It is a collective effort that involves researchers, businesses, and consumers alike, working together to create a marketplace where responsible decisions are both possible and convenient.

\vspace{10mm}

%% file: 7_supplemental_text.tex
\section{Supplementary materials}

\subsection{Aspects}\label{app:aspects}
Table \ref{tab:aspects} gives an overview of the aspects and their description as used in the survey.

\begin{table*}[htb]
\centering
\begin{tabular}{lp{.3\textwidth}p{.6\textwidth}}
\toprule
\textbf{Theme} & \textbf{Aspect} & \textbf{Elaboration} \\
\midrule
A & Price / Value for Money & The cost, whether it's affordable, good value, economical, etc. \\
A & Features / Quality & The specifications, whether it meets your requirements, how well it is made, how good it looks, etc. \\
A & Reviews / Opinions & The opinions and suggestions of others regarding the product, brand, company. \\
B & Brand Appeal / Reputation & The image and impression the brand gives, and whether it is trustworthy, reliable, etc. \\
B & Seller / Retailer & The trust, reliability, reputation, guarantees, etc. of the seller/retailer. \\
B & Governance \& Compliance & The legitimacy and lawfulness of the company, and whether it follows rules/legislation/standards/etc., can be held accountable, is real, etc. \\
C & Ethical Sourcing / Production & How the product was made, and whether the components/materials were sourced ethically, suppliers paid a fair price, etc. \\
C & Slave/Child Labour / Employment Rights & The treatment and rights of workers and whether they are being exploited, the benefits they are provided, etc. \\
C & Country of Origin / Place of Manufacture & The place/country where the products were made/produced. \\
D & Social Impact / Investing in Community & The charity, outreach, support given to communities and organizations and what they are supporting. \\
D & Eco Friendly / Sustainable Practices & The impact that the company has on the environment and ecology (e.g., sustainable growth, carbon neutral, etc.). \\
E & Political Stance / Ideology & The messaging and stance of the company and what it promotes/pushes/believes in. \\
E & Diversity / Equality / Inclusivity & The program and support for DEI. \\
\bottomrule
\multicolumn{3}{p{1\textwidth}}{Aspects were grouped into five themes, not shown to participants: A - Product Evaluation, B - Reputation and Governance, C - Ethical Procurement, D - Environmental and Social Responsibility, E - Inclusion and Ideology. }
\end{tabular}
\caption{\label{tab:aspects} Overview of aspects and their elaboration as shown to participants. }
\end{table*}

\subsection{Survey questions}\label{app:questions}

\newlist{questions}{enumerate}{1}
\setlist*[questions,1]{%
  label=Q\arabic*,
}

\newlist{answers}{enumerate*}{1}
\setlist*[answers,1]{%
  label=\alph*),
}

\paragraph{Part 1: Recent purchase questions}
\begin{questions}

  \item \textbf{Product Description}
    What was the product that you purchased (high-level description, e.g., phone, ice skates, skirt/shorts, etc.)?

  \item \textbf{Product Category}
    What category was the product in?
    \begin{answers}
      \item Clothing, Shoes, Fashion Accessories
      \item Household appliances \& goods, furniture, etc.
      \item Consumer electronics
      \item Books, movies, games, toys
      \item Sports, Recreation, Hobbies, etc.
      \item DIY \& Garden
      \item Other
    \end{answers}

  \item \textbf{Alternatives Considered}
    How many alternatives (different products) did you consider when making your decision?
    \begin{answers}
      \item 1
      \item 2-3
      \item 4-5
      \item 6-9
      \item 10 or more
    \end{answers}

  \item \textbf{Duration}
    Over what period did you research the product and its alternatives?
    \begin{answers}
      \item Over the course of a day
      \item Over the course of a week
      \item Over the course of a month
      \item Longer than a month
    \end{answers}

  \item \textbf{Time Spent}
    How many hours in total did you spend researching the product and its alternatives?
    \begin{answers}
      \item Less than 1 hour
      \item 1-2 hours
      \item 2-4 hours
      \item 5-8 hours
      \item 9-24 hours
      \item 24 hours or more
    \end{answers}

\end{questions}

\paragraph{Part 2: Individual aspect questions}
The following questions pertain to the aspects presented:
    \begin{questions}[resume]
      \item \label{q:importance} \textbf{Importance}
        How important was "[Aspect Name]" when comparing products and their alternatives?
        \begin{answers}
          \item Not at all important
          \item Slightly important
          \item Moderately important
          \item Very important
          \item Extremely important
        \end{answers}

        \item \label{q:considered} \textbf{Considered}
      Did you consider such aspects when comparing products and the brands/companies associated with them?
        \begin{answers}
          \item Yes, I considered it.
          \item No, I didn't consider it.
        \end{answers}

      \item \label{q:searched} \textbf{Search}
        Did you specifically search, look or browse for information about "[Aspect Name]" when comparing products and their alternatives?
        \begin{answers}
          \item Yes, I did (e.g., using a search engine to find reviews, ratings, certifications, etc., browsing product descriptions, reviews, etc.).
          \item No, I did not.
        \end{answers}
\end{questions}

If participants indicated they had searched for information on the presented aspect, they would answer the following questions on 7-point Likert scales:
\begin{questions}[resume]
  \item \label{q:effort} \textbf{Effort}
    \begin{enumerate}
      \item
        I felt that I would have to invest a lot of effort in order to find information about "[Aspect Name]".
        
      \item
        I was willing to invest in finding information about "[Aspect Name]".
        
      \item
        I did not put in a lot of effort to find information about "[Aspect Name]".
    \end{enumerate}
    
  \item \label{q:value} \textbf{Value}
    \begin{enumerate}
      \item
        I thought information about "[Aspect Name]" would be valuable to my purchasing decision.
        
      \item
        I really wanted to know about "[Aspect Name]" as it was key to my decision making.
        
      \item
        I gathered information about "[Aspect Name]" that was useful in making my purchasing decision.
    \end{enumerate}
    
  \item \label{q:success} \textbf{Success expectancies}
    \begin{enumerate}
      \item
        I was not confident that I would be able to find relevant information about "[Aspect Name]".
        
      \item
        I believe I could succeed in finding the relevant information I wanted about "[Aspect Name]".
        
      \item
        I was not successful in finding relevant information about "[Aspect Name]".
    \end{enumerate}
    
  \item \label{q:difficulty} \textbf{Difficulty}
    \begin{enumerate}
      \item
        I knew by searching the internet a lot of information about "[Aspect Name]" can be found.
        
      \item
        I found it difficult to find information about "[Aspect Name]".
        
      \item
        I did not think that it would be possible to find information on "[Aspect Name]".
    \end{enumerate}

\end{questions}

In addition, they would answer the following open-ended question:
\begin{questions}[resume]
  \item \label{q:reasonsyes} \textbf{Reasons}
    Regarding "[Aspect Name]," what information did you want to learn or know about, and why (i.e., how would it have influenced your purchasing decision)?
\end{questions}

If participants indicated they had not searched, they would see the following question instead:
\begin{questions}[resume]
  \item \label{q:reasonsno} \textbf{Reasons}
    If you did not look for information about "[Aspect Name]," what reasons best describe why?
    \begin{answers}
      \item It is not something that I have considered before.
      \item I can never find information about this aspect, so I don't look for it.
      \item It is too time-consuming to find information about this aspect.
      \item There are no good alternatives for this aspect, so I don't look for it.
      \item I already knew the information I needed about this aspect.
      \item It would not have made a difference to my purchasing decision.
      \item I can't do anything about it.
      \item I don't care about it.
      \item Other
    \end{answers}
\end{questions}
The latter option, 'Other', offered a free-text answer possibility. Multiple answers were allowed with this question.

\paragraph{Part 3: Additional insights and EMCB scale}
The survey ended with the following questions:
\begin{questions}[resume]

  \item \textbf{Other Criteria}
    Were there other criteria that you considered to be important? If so, please tell us about them.

  \item \textbf{Additional Information}
    Aside from price, quality, and features, what other information did you seek to learn or find out about the product, company, or seller?

  \item \textbf{Challenges}
    When searching for information online when shopping and comparing products, what are the biggest problems/challenges that you face?

\end{questions}

After these questions, participants answered the \EMCB questionnaire \citep{SUDBURYRILEY20162697}.